\documentclass[10pt,journal]{IEEEtran}

\usepackage{amsmath,amssymb}
\usepackage[dvips]{graphicx}
\usepackage{array}
\usepackage{cite}

\usepackage{color}
\usepackage{float}
\usepackage[mathcal]{euscript}

\usepackage{enumerate}

\usepackage{mathtools}

\usepackage{amsthm}
\newtheorem{proposition}{Proposition}
\newtheorem{lemma}{Lemma}

\newtheorem{theorem}{Theorem}
\newtheorem{example}{Example}

\newtheorem{definition}{Definition}
\newtheorem{property}{Property}

\usepackage{tikz,tabularx}
\usetikzlibrary{decorations.pathreplacing}
\tikzstyle{every picture}+=[remember picture]

\usetikzlibrary{shapes,arrows,shadows}	
\usepgflibrary{patterns}
\usetikzlibrary{patterns}
\usetikzlibrary{shapes.geometric}
\usetikzlibrary{arrows}

\usetikzlibrary{decorations.markings}

\tikzset{myptr1/.style={decoration={markings,mark=at position 1 with %
    {\arrow[scale=2.5]{>}}},postaction={decorate}}}
    
\tikzset{myptr2/.style={decoration={markings,mark=at position 1 with %
    {\arrow[scale=1.8]{>}}},postaction={decorate}}}
    
\def \deg{{\mathrm{deg}}}
\def \Fig {Fig.}

\title{Properties of Syndrome Distribution for Blind Reconstruction of Cyclic Codes}
\author{Arti D.~Yardi and Saravanan Vijayakumaran\\
Department of Electrical Engineering \\
Indian Institute of Technology Bombay, Mumbai 400076, India\\
Email: \{arti,sarva\}@ee.iitb.ac.in}

\begin{document}

\maketitle
\begin{abstract}
In the problem of blind reconstruction of channel codes, the receiver does not have the knowledge of the
channel code used at the transmitter and the aim is to identify this unknown channel code corresponding to the given received sequence.
In this paper, we study this blind reconstruction problem for binary cyclic codes.
In the literature, several researchers have proposed blind reconstruction algorithms that make use of the 
distribution of the syndromes (remainders) of the received polynomials with respect to a candidate polynomial
for the generator polynomial of the cyclic code.
However, very limited analysis is available for the syndrome distribution and its properties.
In this paper, we study the syndrome structure of the received polynomials.
Specifically, we prove that the syndrome distribution of the noise-free sequence can either be uniform or restricted uniform.
We also provide the necessary and sufficient conditions for it to be of the either type.
For the noise-affected received sequence we prove that, finding the syndrome distribution 
is in general computationally intractable.
We also apply these results to analyze the performance of the existing methods and 
verify some of the assumptions made in the literature for blind reconstruction.
\end{abstract}

%
\section{Introduction}
\label{Introduction_structure}

\begin{figure}[t]

\begin{center}
 
    \begin{tikzpicture}

	  \node [left] at (0.6,0) {Data};
	  \draw [-,thick] (0.6,0) -- (1,0);
	  \draw [myptr1] (0.6,0) -- (1,0);	  

	  \draw [rounded corners, thick] (1,-0.55) rectangle (3.3,0.5);  
	  
	  \node [right] at (1.3,0.25) {Unknown};
	  \node [right] at (1,-0.25) {Channel Code};

	  \draw [-,thick] (3.3,0) -- (3.3+0.8,0);	  
	  \draw [myptr1] (3.3,0) -- (3.3+0.8,0);	  	  

	  \draw (3.3+0.8+0.15,0)[thick] circle (0.15cm);
	  \draw [-,thick] (3.3+0.8+0.15-0.07,0) -- (3.3+0.8+0.15+0.07,0);
	  \draw [-,thick] (3.3+0.8+0.15,0.07) -- (3.3+0.8+0.15,-0.07);

  	  \draw [-,thick] (3.3+0.8+0.15,0.15) -- (3.3+0.8+0.15,0.8);
  	  \draw [myptr1] (3.3+0.8+0.15,0.8) -- (3.3+0.8+0.15,0.15);  	  
  	  
	  \node [above] at (3.3+0.8+0.15,0.8) {Noise};	  
	  
	  \draw [-,thick] (3.3+0.8+0.3,0) -- (3.3+0.8+0.3+0.8,0);
	  \draw [myptr1] (3.3+0.8+0.3,0) -- (3.3+0.8+0.3+0.8,0);	  	  
	  
	  \draw [rounded corners,thick] (3.3+0.8+0.3+0.8,-0.55) rectangle (3.3+0.8+0.3+0.8+3.2,0.5);  
	  
	  \node [right] at (3.3+0.8+0.3+0.8+0.05,0.25) {Blind reconstruction};
	  \node [right] at (3.3+0.8+0.3+0.8+0.25,-0.25) {of channel code};	  
	  
	  \draw [-,thick] (3.3+0.8+0.3+0.8+1.6,-0.55) -- (3.3+0.8+0.3+0.8+1.6,-0.55-0.6);
	  \draw [myptr1] (3.3+0.8+0.3+0.8+1.6,-0.55) -- (3.3+0.8+0.3+0.8+1.6,-0.55-0.6);	  	  	  

	  \draw [rounded corners,thick] (3.3+0.8+0.3+0.8+0.35,-0.55-0.6-1) rectangle (3.3+0.8+0.3+0.8+3.2-0.35,-0.55-0.6);  	  
	  
	  \node [right] at (3.3+0.8+0.3+0.8+0.4,0.25-1.65) {Reconstructed};
	  \node [right] at (3.3+0.8+0.3+0.8+0.5,-0.25-1.6) {channel code};	  

	  \draw [-,thick] (3.3+0.8+0.3+0.8+1.6,-0.55-0.6-1) -- (3.3+0.8+0.3+0.8+1.6,-0.55-0.6-1-0.4);
	  \draw [myptr1] (3.3+0.8+0.3+0.8+1.6,-0.55-0.6-1) -- (3.3+0.8+0.3+0.8+1.6,-0.55-0.6-1-0.4);	  	  	  	  
	  
	  \node [below] at (3.3+0.8+0.3+0.8+1.6,-0.55-0.6-1-0.4) {Decode data};	  	  
	  
	  \node [right] at (1.3,-1.75) {Noise-affected};	  
	  \node [right] at (1.5,-1.7-0.5) {codewords};	  	  
	  \node [right] at (1.15,-1.7-0.5-0.5) {(Received data)};	  	  	  
	  
	  \draw [-,dashed,thick] (3.3+0.8+0.3+0.4,-0.1) -- (3.3+0.8+0.3+0.4,-0.5);	  	  	  	  	  
	  \draw [myptr2] (3.3+0.8+0.3+0.4,-0.5) -- (3.3+0.8+0.3+0.4,-0.1);	  	  	  	  	  
	  
	  \draw [-,dashed,thick] (2.5,-1.5) -- (3.3+0.8+0.3+0.4,-0.5);	  	  	  	  	  	  

    \end{tikzpicture}
\end{center}

\caption{A system model for blind reconstruction problem of channel codes.}
\label{Figure_blind_reconstruction_problem_intro}
\end{figure}
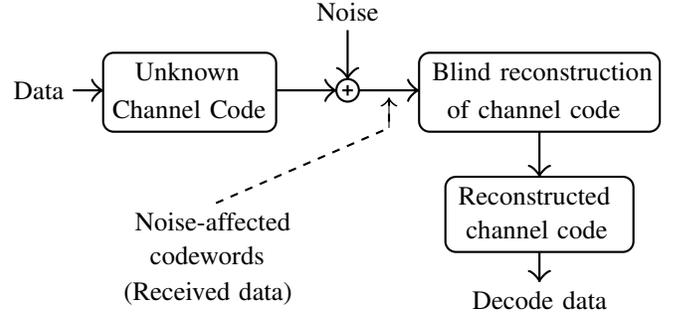

Channel codes play a vital role in the digital communication system to make the system robust to the errors introduced by the channel noise.
When the channel code used at the transmitter is known at the receiver, the received data can be decoded to obtain the 
transmitted messages~\cite{LinCostello2004}.
However there could be situations when the channel code used at the transmitter is not known at the receiver.
For example, in military surveillance the channel code used by an adversary might not be known. 
In such scenarios, in order to decode the received data, one needs to first identify this unknown 
channel (see \Fig~\ref{Figure_blind_reconstruction_problem_intro}).
This problem of identifying the channel corresponding to the given received data is known as
\textit{blind reconstruction of channel codes}~\cite{Rice95, Planquette96, Filiol97}.
%

This blind reconstruction problem is in general NP-hard~\cite{Valembois2001}.
While identifying a particular channel code, it is typically assumed that the family of the code, such as convolutional or 
linear block code, is known. 
The underlying structure of this particular family is then used to identify the code.
In the literature, various algorithms have been proposed for blind reconstruction of
convolutional codes~\cite{Marazin2011,DingelHau2007}, turbo codes~\cite{Barbier2005,CoteSen2010}, 
linear block codes~\cite{Valembois2001,SicotHouBaJournal2009,CluF2009}, LDPC codes~\cite{Cluzeau2006,Moosavi_journal}, 
and cyclic codes~\cite{LeeSong2012_Korea,Chabot_thesis,EuropeanWireless2014, TCOMM_2016, Zhou2013_Entropy_new,Zhou2013_Entropy}.
%

%

Chabot~\cite{Chabot_thesis}, Lee et al.~\cite{LeeSong2012_Korea}, and Yardi et al.~\cite{EuropeanWireless2014}
have studied this blind reconstruction problem for cyclic codes when the length of the code is assumed to be known at the receiver.
Zhou et al.~\cite{Zhou2013_Entropy_new, Zhou2013_Entropy} and Yardi et al.~\cite{TCOMM_2016} consider the situation
when the length of the cyclic code is not known.
In this paper, we focus on the unknown length scenario.
For the unknown length scenario, a key idea proposed in the existing methods is summarized next~\cite{TCOMM_2016, Zhou2013_Entropy_new, Zhou2013_Entropy}.
The unknown cyclic code $C(n_0,g_0)$ is identified by finding its length $n_0$ and the factors of its generator polynomial $g_0(X)$.
Since the first received bit might not be the first bit of a received codeword, for blind reconstruction, 
one also needs to identify the location of the codeword boundaries or synchronization of the received data.
The analysis begins by assuming a length $n$,
synchronization, and a candidate polynomial $f(X)$ for the factor of the generator polynomial.
Note that $f(X)$ is factor of $X^n+1$ since for an assumed $n$, the generator polynomial has to be a factor of $X^n+1$~\cite{LinCostello2004}.
For the assumed $n$, synchronization, and $f(X)$ there are the following two cases.
\begin{enumerate}[(a)]
\item Both $n$ and synchronization are correct, and $f(X)$ is a factor of $g_0(X)$
\item Either $n$ or synchronization is not correct or $f(X)$ is not a factor of $g_0(X)$
\end{enumerate}
For the chosen $n$, synchronization, and $f(X)$, the key step in the existing methods
consists of determining which one of the above two cases holds.

In order to use the optimal likelihood ratio tests for determining whether (a) is true or (b) is true, one needs to 
find the probability of the received data when condition (a) is true and when condition (b) is true~\cite{Poor94}.
However, we next explain that finding this probability is, in general, computationally intractable.
When condition (a) is true, let $\mathbb{P}[\mathbf{y}]$ be the probability of receiving an $n_0$-bit vector $\mathbf{y}$.
This probability can be computed by conditioning over all possible transmitted codewords in $C(n_0,g_0)$ as follows,
\begin{align*}
\mathbb{P}[\mathbf{y}] = \sum_{\mathbf{v} \in C(n_0,g_0)} \mathbb{P}\Big[\mathbf{y} \Big| \mathbf{v} \mbox{ is transmitted}\Big]\mathbb{P}\Big[\mathbf{v} \mbox{ is transmitted}\Big].
%
\end{align*}
When $f(X)=g_0(X)$ and the true code $C(n_0,g_0)$ used at the transmitter is known at the receiver, it is shown in \cite{ISIT2014} that
finding $\mathbb{P}[\mathbf{y}]$ is, in general, computationally intractable.
Since in our case $C(n_0,g_0)$ is not known, obtaining $\mathbb{P}[\mathbf{y}]$ is even more computationally 
intractable.
Hence in the literature, researchers have proposed suboptimal tests which make use of the syndromes of the received polynomials 
to take a decision between (a) and (b)~\cite{TCOMM_2016, Zhou2013_Entropy_new, Zhou2013_Entropy}.
In \cite{TCOMM_2016} and \cite{Zhou2013_Entropy}, the properties of the zero syndromes of the received polynomials are used
to distinguish between (a) and (b). Whereas in \cite{Zhou2013_Entropy_new}, the marginal distribution of
the coefficients of the syndromes is used for blind reconstruction.

Understanding the syndrome structure of the received polynomials is thus important to
study the problem of blind reconstruction of binary cyclic codes.
However, very limited analysis is available for the syndrome distribution and its properties.
Due to lack of knowledge of the syndrome distribution, typically in the literature some assumptions are made
to simplify the analysis~\cite{Zhou2013_Entropy,Zhou2013_Entropy_new}. 
For example, in \cite{Zhou2013_Entropy_new} it is assumed that when either of the assumed parameter is incorrect (case (b) mentioned above), 
every coefficient in the syndrome of the received polynomial is equally likely to be zero or one.
In \cite{Zhou2013_Entropy}, the received data is assumed to behave as a random bitstream for the incorrect parameters.
In this paper, we analyze the properties of the syndrome distribution and verify these assumptions.
These syndrome properties can also be use to study the theoretical performance of the method proposed in \cite{TCOMM_2016}.
The main contributions of this paper are as follows.
\begin{enumerate}[(1)]
\item We first characterize the syndrome distribution of the noise-free polynomials with respect to a candidate polynomial $f(X)$.
We prove that when either of the assumed parameter are incorrect (case (b) mentioned above), the distribution of the syndrome can 
be either uniform or restricted uniform (see (\ref{Eqn_type_2_distri}), (\ref{Eqn_type_3_distri}), 
and Proposition~\ref{Proposition_cyclic_structure_no_type1}).
We also provide the necessary and sufficient conditions for the distribution to be restricted uniform 
(see Theorems~\ref{Theorem_main_iff_structure}, \ref{Theorem_W_suff_pre_cyclic_structure}, and \ref{Theorem_structure_w_prime_type3}).
\item We study the syndrome distribution of the noise-affected received polynomials. 
We prove that when the syndrome distribution of the noise-free polynomial
is uniform, the distribution of the noise-affected polynomial would also be uniform (see Theorem~\ref{Theorem_wj_yj_equally_likely}).
We also show that, when the distribution of the syndrome of the noise-free polynomial is restricted uniform, 
finding the distribution of the noise-affected polynomial is in general computationally intractable.
\item Finally, using the syndrome analysis mentioned in (1) and (2) above,
we verify the assumptions made in \cite{Zhou2013_Entropy} and \cite{Zhou2013_Entropy_new} and provide a 
theoretical analysis of the blind reconstruction method proposed in \cite{TCOMM_2016}.
\end{enumerate}
%

\textit{Organization:}
The system model for blind reconstruction of cyclic codes and some preliminaries are provided in Section~\ref{Section_System_model}.
We study the syndrome distribution of the noise-free sequence in Section~\ref{Section_cyclic_structure_noise_free}.
This analysis is then extended to the noise-affected case in Section~\ref{Section_cyclic_structure_noise_affected}.
In Section~\ref{Section_Blind_Reconstruction_cyclic_structure}, we provide a theoretical analysis of the existing blind reconstruction methods.
Finally, we conclude in Section~\ref{Section_conclusion}.

\textit{Notation:} 
The set of natural numbers is denoted by $\mathbb{N}$ and $\mathbb{F}_2$ denotes the finite field with two elements $0$ and $1$.
The polynomial ring with coefficients from $\mathbb{F}_2$ is denoted by $\mathbb{F}_2[X]$.
The integer $\lfloor m \rfloor$ denotes the greatest integer less than or equal to $m$.
We use boldface letters to denote the vectors and lower case letters for the components of a vector. 
For example, vector $\mathbf{y} = \begin{bmatrix} y_{0} & y_1 & \ldots & y_{n-1} \end{bmatrix}$, where $y_i$ for $i = 0, 1, \ldots, n-1$
are the components of $\mathbf{y}$.
The polynomial representation of vector $\mathbf{y}$, is given by $\mathbf{y}(X) = y_{0} + y_{1}X + \ldots + y_{n-1}X^{n-1}$.
Note that the polynomials corresponding to vectors are denoted by boldface letters.
For integers $l,r$, $0 \leq l < r < n$, we define $\mathbf{y}(l:r) \coloneqq \begin{bmatrix} y_{l} & y_{l+1} & \ldots & y_{r} \end{bmatrix}$.
When $l=0$, the vector $\mathbf{y}(0:r)$ is called as a \textit{prefix} of $\mathbf{y}$ and 
when $r=n-1$, the vector $\mathbf{y}(l:n-1)$ is called as a \textit{suffix} of $\mathbf{y}$.

\section{System model and preliminaries} 
\label{Section_System_model}

A linear block code of length $n$ is denoted by $C(n)$ and the cyclic code of length $n$ and the generator polynomial $g(X)$ is denoted by 
$C(n,g)$. 
Let $k$ be the dimension of $C(n,g)$.
It is known that $k = n - \deg(g)$, where $\deg(g)$ is the degree of $g(X)$~\cite{LinCostello2004}.
When $k=0$ or $k=n$, the code $C(n,g)$ is said to be a trivial cyclic code.
%
%
Any codeword polynomial $\mathbf{v}(X)$ can be written as $\mathbf{v}(X) = \mathbf{u}(X)g(X)$
where $\mathbf{u}(X)$ is a message polynomial.
%
The set of polynomials in $\mathbb{F}_2[X]$ of degrees strictly less than $n$ is denoted by $\mathcal{P}_n$, i.e., 
\begin{align}
\mathcal{P}_n = \left\{f(X) \in \mathbb{F}_2[X] \Big| \mathrm{deg}(f(X)) \leq n-1\right\}.
\end{align}
Using to this notation, $\mathbf{v}(X) \in \mathcal{P}_{n}$ and $\mathbf{u}(X) \in \mathcal{P}_{k}$.

Suppose the cyclic code $C(n_0,g_0)$ of dimension $k_0$ is used at the transmitter.
Each transmitted codeword is independent and identically distributed (i.i.d.) according to the uniform distribution
over the set of codewords of $C(n_0,g_0)$.
We assume that the noise is introduced by a binary symmetric channel (BSC) of crossover probability $p < 1/2$. 
The received bitstream is denoted by $y_0, y_1, \ldots , y_{N-1}$.
We now define the synchronization $s_0$ of this bitstream as follows.

\begin{definition}
\label{Definition_sync}
%
The synchronization $s_0$ of the received bitstream $y_0, y_1, \ldots , y_{N-1}$ is defined as the 
smallest integer such that the vector $\begin{bmatrix} y_{s_0} & \ldots & y_{s_0+n_0-1} \end{bmatrix}$ of length $n_0$
is the noise-affected version of the transmitted codeword of the cyclic code $C(n_0,g_0)$ used at the transmitter. 
Note that $0 \leq s_0 < n_0$.
\hfill $\square$
\end{definition}
%

%
Let $n \in \mathbb{N}$ be an assumed length of the code.
For an assumed synchronization $s$, $0 \leq s < n$ ignore $y_0, y_1, \ldots , y_{s-1}$
from the received bitstream and divide the remaining bitstream into vectors of length $n$.
Thus the first $n$-bit vector is given by $\mathbf{y}_{1}(n,s) = \begin{bmatrix} y_s & \ldots & y_{s+n-1} \end{bmatrix}$.
Similarly the $j$th $n$-bit vector is given by $\mathbf{y}_{j}(n,s) = \begin{bmatrix} y_{s+(j-1)n} & \ldots & y_{s+jn-1}\end{bmatrix}$.
Suppose we have received $M = \lfloor (N-s)/n \rfloor$ vectors of length $n$. 
For the sake of simplicity we will drop parameters $n$ and $s$ from $\mathbf{y}_{j}(n,s)$.
Thus $\mathbf{y}_1, \mathbf{y}_2, \ldots, \mathbf{y}_M$ is the sequence of $n$-bit vectors
for an assumed synchronization $s$.
Note that, the polynomial corresponding to $\mathbf{y}_j$ is given by $\mathbf{y}_j(X)$, for $j = 1,2,\ldots, M$.

We assume that cyclic code $C(n_0,g_0)$ used at the transmitter is non-degenerate, where 
a degenerate code is defined as follows.
\begin{definition}
\label{Definition_degerate}
Let $G$ be a generator matrix of a linear block code $C(n)$.
Then $C(n)$ is said to be degenerate if $G$ can be written as,
\begin{align}
G =  \Big[\underbrace{G^{\prime} \mbox{~~} G^{\prime} \mbox{~~} \cdots \mbox{~~} G^{\prime}}_{l \text{ times}}\Big],
\label{Eqn_Definition_self_repetition_family}
\end{align}
where $l\in \mathbb{N}, l > 1$ and $G^{\prime}$ is a generator matrix of some other linear block code $C^{\prime}(n^{\prime})$
of length $n^{\prime} = n/l$~\cite[Ch.~8]{Macwilliams_Sloane_1977}.
The code $C^{\prime}(n^{\prime})$ is said to be a component code of $C(n)$.
For a cyclic code, its component code is also cyclic~\cite{TCOMM_2016}.
\hfill $\square$
\end{definition}
For blind reconstruction of a degenerate cyclic code, it is sufficient to identify its 
non-degenerate component (see \cite{TCOMM_2016} for details).
Hence without loss of generality we consider the situation when the cyclic code used at the transmitter is not degenerate.

\subsection{Preliminaries}
In this section, we consider some preliminaries that will be required throughout the paper.

%
\begin{definition}
\label{Definition_order_fX}
The order of a polynomial $f(X)$ over $\mathbb{F}_2(X)$ is defined as the least positive integer $l$ such that 
$f(X)$ divides $X^{l}+1$~\cite[Sec.~3.1]{Lidl86}.
\hfill $\square$
\end{definition}

We next recall a definition of a \textit{linear recurring sequence} and its \textit{period}.
\begin{definition}
\label{Definition_LRR}
For a positive integer $l$, a sequence of bits $v_0, v_1, \cdots $ is said to be a linear recurring sequence of $l^{th}$ order 
if they follow a relation
\begin{align}
%
v_{r+l} = \sum_{i=0}^{l-1} h_{i} v_{r+i}, \mbox{~~for~} r = 0,1,\ldots
\label{Eqn_LRR_defi}
\end{align}
where $h_i \in \mathbb{F}_2$ for $i = 0,1,\ldots,l-1$.
It is known that any linear recurring sequence is ultimately periodic and its period is 
defined as a positive integer $n$ such that $v_{r+n} = v_{r}$, for $r = 0,1,\ldots$~\cite[Sec.~6.1]{Lidl86}.
\hfill $\square$
\end{definition}

The \textit{minimal polynomial} associated with a linear recurring sequence is defined next.
\begin{definition}
\label{Definition_min_poly_LRR}
Suppose $d$ is the least positive integer such that the linear recurring sequence $v_0, v_1, \cdots $ 
satisfies the relation given in (\ref{Eqn_LRR_defi}).
Then the polynomial $h(X) \coloneqq h_{d-1} X^{d-1} + h_{d-2} X^{d-2} + \ldots + h_{0}$ is called as
the minimal polynomial associated with this sequence~\cite[Sec.~6.4]{Lidl86}. 
\hfill $\square$
\end{definition}
It is known that any linear recurring sequence has a unique minimal polynomial
and the order of the minimal polynomial is equal to the least period of this sequence ~\cite[Sec.~6.4]{Lidl86}.
We next define a vector of \textit{degenerate pattern}.
\begin{definition}
\label{Definition_degenerate_pattern}
An $n$-bit vector $\mathbf{v}$ is said to be of \textit{degenerate pattern} if it can be written as 
\begin{align}
\mathbf{v} = \Big[\underbrace{\mathbf{w} \mbox{~~} \mathbf{w} \mbox{~~} \cdots \mbox{~~} \mathbf{w}}_{l \text{ times}}\Big],
\label{Eqn_Definition_degenerate_pattern}
\end{align}
where $l \in \mathbb{N}$ and $\mathbf{w}$ itself is a vector of length $n^{\prime} = n/l$ such that
$\mathbf{w}$ is not a vector of degenerate pattern~\cite{Cancellieri_2015}.
\hfill $\square$
\end{definition}

It is known that the sequence of bits given by $[\mathbf{w} \mbox{~~} \mathbf{w} \mbox{~~} \cdots]$ in (\ref{Eqn_Definition_degenerate_pattern})
is a linear recurring sequence with least period $n^{\prime}$~\cite{Lidl86}.
We now define the \textit{minimal generating polynomial} associated with this sequence as follows.

%
\begin{definition}
\label{Definition_min_gen_poly_LRR}
Let $h(X)$ be the minimal polynomial of the linear recurring sequence
$[\mathbf{w} \mbox{~~} \mathbf{w} \mbox{~~} \cdots]$ with least period $n^{\prime}$.
Then the \textit{minimal generating polynomial} $m(X)$ associated with
$[\mathbf{w} \mbox{~~} \mathbf{w} \mbox{~~} \cdots]$ is defined as 
\begin{align}
m(X) \coloneqq \frac{X^{n^{\prime}} + 1}{h^{\prime}(X)},
\end{align}
where $h^{\prime}(X) = X^{\deg(h)} h(X^{-1})$.
It is known that the polynomial $\mathbf{w}(X)$ corresponding to $\mathbf{w}$ is a multiple of $m(X)$~\cite[Sec.~7.4]{Peterson_1996}.
Note that $h(X)$ is the generator polynomial of the dual code of $C(n^{\prime},m)$.
\hfill $\square$
\end{definition}

We next provide a definition of the \textit{outer direct sum} of two linear block codes.
\begin{definition}
\label{Definition_outer_direct_sum} 
The outer direct sum $C_1(n_1) + C_2(n_2)$ of codes $C_1(n_1)$ and $C_2(n_2)$ 
is defined as a linear block code formed by concatenating all possible codewords of $C_1(n_1)$ with 
all possible codewords of $C_2(n_2)$, i.e.,
\begin{align*}
C_1(n_1) + C_2(n_2) \coloneqq \left\{\begin{bmatrix} \mathbf{v}  & \mathbf{w} \end{bmatrix} \bigg| \mathbf{v} \in C_1(n_1), \mathbf{w} \in C_2(n_2) \right\}. 
%
\end{align*}
\hfill $\square$
\end{definition}
%

\begin{figure}[t]

\begin{center}
 
    \begin{tikzpicture}[scale=0.77]

	  \draw [-][thick] (-0.3,0) -- (2.8,0);
	  \draw [black, fill=black](0,1.5) circle (0.05cm);
	  
	  \draw [black, fill=black](0,0) circle (0.025cm);	  
	  \draw [black, fill=black](0.35,0) circle (0.025cm);
	  \draw [black, fill=black](0.35+0.35,0) circle (0.025cm);
	  \draw [black, fill=black](0.35+0.35+0.35,0) circle (0.025cm);
	  \draw [black, fill=black](0.35+0.35+0.35+0.35,0) circle (0.025cm);
	  \draw [black, fill=black](0.35+0.35+0.35+0.35+0.35,0) circle (0.025cm);
	  \draw [black, fill=black](0.35+0.35+0.35+0.35+0.35+0.35,0) circle (0.025cm);
	  \draw [black, fill=black](0.35+0.35+0.35+0.35+0.35+0.35+0.35,0) circle (0.025cm);

	  \draw [-,ultra thick] (0,0) -- (0,1.5);

	  \node [below] at (0,0) {\footnotesize{$0$}};
	  \node [below] at (0.35,0) {\footnotesize{$1$}};
	  \node [below] at (0.35+0.35,0) {\footnotesize{$2$}};
	  \node [below] at (0.35+0.35+0.35,0) {\footnotesize{$3$}};
	  \node [below] at (0.35+0.35+0.35+0.35,0) {\footnotesize{$4$}};
	  \node [below] at (0.35+0.35+0.35+0.35+0.35,0) {\footnotesize{$5$}};
	  \node [below] at (0.35+0.35+0.35+0.35+0.35+0.35,0) {\footnotesize{$6$}};
	  \node [below] at (0.35+0.35+0.35+0.35+0.35+0.35+0.35,0) {\footnotesize{$7$}};

	  \node [left] at (0,1.5) {\footnotesize $1$};

	  \node [below] at (1.3,-1) {(a) Degenerate};

	  \draw [-] [thick] (-0.5+4-0.15,0) -- (2.8+4-0.15,0);
	  
	  \draw [black, fill=black](0+4-0.15,0) circle (0.025cm);	  
	  \draw [black, fill=black](0.35+4-0.15,0) circle (0.025cm);
	  \draw [black, fill=black](0.35+0.35+4-0.15,0) circle (0.025cm);
	  \draw [black, fill=black](0.35+0.35+0.35+4-0.15,0) circle (0.025cm);
	  \draw [black, fill=black](0.35+0.35+0.35+0.35+4-0.15,0) circle (0.025cm);
	  \draw [black, fill=black](0.35+0.35+0.35+0.35+0.35+4-0.15,0) circle (0.025cm);
	  \draw [black, fill=black](0.35+0.35+0.35+0.35+0.35+0.35+4-0.15,0) circle (0.025cm);
	  \draw [black, fill=black](0.35+0.35+0.35+0.35+0.35+0.35+0.35+4-0.15,0) circle (0.025cm);
	  
	  \draw [black, fill=black](0+4-0.15,0.4) circle (0.05cm);
	  \draw [black, fill=black](0.35+4-0.15,0.4) circle (0.05cm);
	  \draw [black, fill=black](0.35+0.35+4-0.15,0.4) circle (0.05cm);
	  \draw [black, fill=black](0.35+0.35+0.35+4-0.15,0.4) circle (0.05cm);
	  \draw [black, fill=black](0.35+0.35+0.35+0.35+4-0.15,0.4) circle (0.05cm);
	  \draw [black, fill=black](0.35+0.35+0.35+0.35+0.35+4-0.15,0.4) circle (0.05cm);
	  \draw [black, fill=black](0.35+0.35+0.35+0.35+0.35+0.35+4-0.15,0.4) circle (0.05cm);
	  \draw [black, fill=black](0.35+0.35+0.35+0.35+0.35+0.35+0.35+4-0.15,0.4) circle (0.05cm);
	  
	  \draw [-,ultra thick] (0+4-0.15,0) -- (0+4-0.15,0.4);
	  \draw [-,ultra thick] (0.35+4-0.15,0) -- (0.35+4-0.15,0.4);
	  \draw [-,ultra thick] (0.35+4+0.35-0.15,0) -- (0.35+4+0.35-0.15,0.4);
	  \draw [-,ultra thick] (0.35+4+0.35+0.35-0.15,0) -- (0.35+4+0.35+0.35-0.15,0.4);
	  \draw [-,ultra thick] (0.35+4+0.35+0.35+0.35-0.15,0) -- (0.35+4+0.35+0.35+0.35-0.15,0.4);
	  \draw [-,ultra thick] (0.35+4+0.35+0.35+0.35+0.35-0.15,0) -- (0.35+4+0.35+0.35+0.35+0.35-0.15,0.4);
	  \draw [-,ultra thick] (0.35+4+0.35+0.35+0.35+0.35+0.35-0.15,0) -- (0.35+4+0.35+0.35+0.35+0.35+0.35-0.15,0.4);
	  \draw [-,ultra thick] (0.35+4+0.35+0.35+0.35+0.35+0.35+0.35-0.15,0) -- (0.35+4+0.35+0.35+0.35+0.35+0.35+0.35-0.15,0.4);

	  \node [below] at (0+4-0.15,0) {\footnotesize{$0$}};
	  \node [below] at (0.35+4-0.15,0) {\footnotesize{$1$}};
	  \node [below] at (0.35+0.35+4-0.15,0) {\footnotesize{$2$}};
	  \node [below] at (0.35+0.35+0.35+4-0.15,0) {\footnotesize{$3$}};
	  \node [below] at (0.35+0.35+0.35+0.35+4-0.15,0) {\footnotesize{$4$}};
	  \node [below] at (0.35+0.35+0.35+0.35+0.35+4-0.15,0) {\footnotesize{$5$}};
	  \node [below] at (0.35+0.35+0.35+0.35+0.35+0.35+4-0.15,0) {\footnotesize{$6$}};
	  \node [below] at (0.35+0.35+0.35+0.35+0.35+0.35+0.35+4-0.15,0) {\footnotesize{$7$}};
	  
	  \draw [-,densely dotted] (0+4-0.15,0.4) -- (0.35+0.35+0.35+0.35+0.35+0.35+0.35+4-0.15,0.4);
	  \node [left] at (4-0.15,0.4) {\footnotesize $0.125$};

 	  \node [below] at (2+2.9-0.15,-1) {(b) Uniform};
 
	  \draw [-] [thick] (-0.5+4+4-0.15,0) -- (3+4+4-0.15,0);
	  
	  \draw [black, fill=black](0+4+4-0.15,0) circle (0.025cm);	  
	  \draw [black, fill=black](0.35+4+4-0.15,0) circle (0.025cm);
	  \draw [black, fill=black](0.35+0.35+4+4-0.15,0) circle (0.025cm);
	  \draw [black, fill=black](0.35+0.35+0.35+4+4-0.15,0) circle (0.025cm);
	  \draw [black, fill=black](0.35+0.35+0.35+0.35+4+4-0.15,0) circle (0.025cm);
	  \draw [black, fill=black](0.35+0.35+0.35+0.35+0.35+4+4-0.15,0) circle (0.025cm);
	  \draw [black, fill=black](0.35+0.35+0.35+0.35+0.35+0.35+4+4-0.15,0) circle (0.025cm);
	  \draw [black, fill=black](0.35+0.35+0.35+0.35+0.35+0.35+0.35+4+4-0.15,0) circle (0.025cm);
	  
	  \draw [black, fill=black](0+4+4-0.15,0.8) circle (0.05cm);
	  \draw [black, fill=black](0.35+0.35+4+4-0.15,0.8) circle (0.05cm);
	  \draw [black, fill=black](0.35+0.35+0.35+4+4-0.15,0.8) circle (0.05cm);
	  \draw [black, fill=black](0.35+0.35+0.35+0.35+0.35+0.35+0.35+4+4-0.15,0.8) circle (0.05cm);

	  \draw [-,ultra thick] (0+4+4-0.15,0) -- (0+4+4-0.15,0.8);
	  \draw [-,ultra thick] (0.35+4+0.35+4-0.15,0) -- (0.35+4+0.35+4-0.15,0.8);
	  \draw [-,ultra thick] (0.35+4+0.35+0.35+4-0.15,0) -- (0.35+4+0.35+0.35+4-0.15,0.8);
	  \draw [-,ultra thick] (0.35+4+0.35+0.35+0.35+0.35+0.35+0.35+4-0.15,0) -- (0.35+4+0.35+0.35+0.35+0.35+0.35+0.35+4-0.15,0.8);

	  \node [below] at (0+4+4-0.15,0) {\footnotesize{$0$}};
	  \node [below] at (0.35+4+4-0.15,0) {\footnotesize{$1$}};
	  \node [below] at (0.35+0.35+4+4-0.15,0) {\footnotesize{$2$}};
	  \node [below] at (0.35+0.35+0.35+4+4-0.15,0) {\footnotesize{$3$}};
	  \node [below] at (0.35+0.35+0.35+0.35+4+4-0.15,0) {\footnotesize{$4$}};
	  \node [below] at (0.35+0.35+0.35+0.35+0.35+4+4-0.15,0) {\footnotesize{$5$}};
	  \node [below] at (0.35+0.35+0.35+0.35+0.35+0.35+4+4-0.15,0) {\footnotesize{$6$}};
	  \node [below] at (0.35+0.35+0.35+0.35+0.35+0.35+0.35+4+4-0.15,0) {\footnotesize{$7$}};
	  
	  \draw [-,densely dotted] (0+4+4-0.15,0.8) -- (0.35+0.35+0.35+0.35+0.35+0.35+0.35+4+4-0.15,0.8);
	  \node [left] at (4+4-0.15,0.8) {\footnotesize $0.25$};

 	  \node [below] at (2+3.1+4-0.15,-1) {(c) Restricted uniform};

    \end{tikzpicture}
\end{center}

\caption{An illustration of the three types of distributions defined in (\ref{Eqn_type_1_distri}), (\ref{Eqn_type_2_distri}),
and (\ref{Eqn_type_3_distri}) for a random variable $X$ with the support set $\mathcal{X}=\{0,1,\ldots,7\}$.}
\label{Figure_example_distributions}
\end{figure}
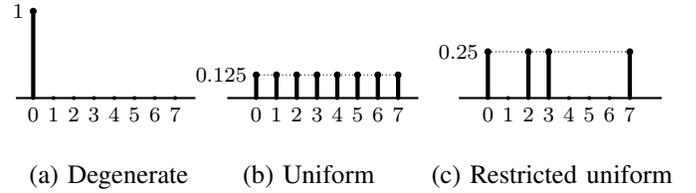

We now define three types of distributions for a discrete random variable $X$ with a finite support set $\mathcal{X}$
such that the cardinality of $|\mathcal{X}|$ of set $\mathcal{X}$ is equal to $2^L$ for some integer $L \geq 1$.
An example situation for these three types of distributions is shown in \Fig~\ref{Figure_example_distributions}.
%
%
%
%
\begin{enumerate}[(1)]
\item Degenerate distribution (see \Fig~\ref{Figure_example_distributions}(a))\\
Random variable $X$ is said to follow the degenerate distribution if it takes a particular value $x_0 \in \mathcal{X}$ 
with probability one, i.e., 
\begin{align}
\mathbb{P}[X = x] = \left\{
			\begin{array}{l l}
			    1 & \quad \text{if } x = x_0 \text{ for some }x_0 \in \mathcal{X}, \\
			    0 & \quad \text{otherwise}.
			\end{array} 
		  \right.
\label{Eqn_type_1_distri}
\end{align}
\item Uniform distribution (see \Fig~\ref{Figure_example_distributions}(b)) \\
When random variable $X$ follows the uniform distribution on its support set $\mathcal{X}$, 
\begin{align}
\mathbb{P}[X = x] = \left\{
			\begin{array}{l l}
			    1/|\mathcal{X}| & \quad \text{if } x \in \mathcal{X}, \\
			    0 & \quad \text{otherwise}.
			\end{array} 
		  \right.
\label{Eqn_type_2_distri}		  
\end{align}

\item Restricted uniform distribution (see \Fig~\ref{Figure_example_distributions}(c)) \\
Consider a strict subset $\mathcal{X}_0$ of $\mathcal{X}$ such that $|\mathcal{X}_0| = 2^l$ for some integer $l$, where $1 \leq l < L$.
Random variable $X$ is said to follow the restricted uniform distribution on $\mathcal{X}$ if it follows 
the uniform distribution on set $\mathcal{X}_0$, i.e., 
\begin{align}
\mathbb{P}[X = x] = \left\{
			\begin{array}{l l}
			    1/|\mathcal{X}_0| & \quad \text{if } x \in \mathcal{X}_0, \\
			    0 & \quad \text{otherwise}.
			\end{array} 
		  \right.
\label{Eqn_type_3_distri}		  
\end{align}
\end{enumerate}
%


\section{Syndrome distribution of the noise-free sequence}
\label{Section_cyclic_structure_noise_free}

Recall that for an assumed length $n$ and synchronization $s$, the received sequence of polynomials is given by
$\mathbf{y}_1(X),\mathbf{y}_2(X),\ldots,\mathbf{y}_M(X)$ (see Section~\ref{Section_System_model}).
Suppose $f(X)$ is factor of $X^n+1$.
For blind reconstruction, we need to study the distribution of $\mathbf{y}_j(X) \bmod f(X)$, for $j = 1,2,\ldots, M$.
Suppose the $j$th received polynomial $\mathbf{y}_j(X)$ is given by,
\begin{align}
\mathbf{y}_j(X) = \mathbf{w}_j(X) + \mathbf{e}_j(X),
\label{Equation_yj_wj_ej_temp}
\end{align}
where $\mathbf{w}_j(X)$ is the noise-free polynomial and $\mathbf{e}_j(X)$ the error polynomial.
In this section, we study the distribution of the syndrome of the noise-free polynomial, i.e., the distribution of $\mathbf{w}_j(X) \bmod f(X)$, 
where $1 \leq j \leq M$.
The distribution of $\mathbf{y}_j(X) \bmod f(X)$ will be studied in the next section.
%

%
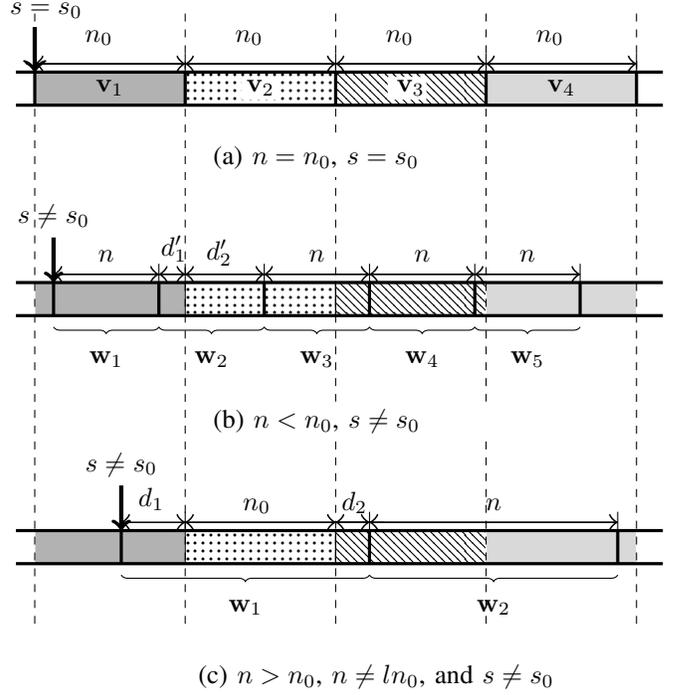
\begin{figure}[t]

\begin{center}
 
    \begin{tikzpicture}


	  \draw [white,fill=black!30] (0.15,0+3.1) rectangle (0.15+2,0.45+3.1);
	  \draw [white,fill=black!30] (0.15,0+3.3-3) rectangle (0.15+2,0.45+3.3-3);
	  \draw [white,fill=black!30] (0.15,0+3.3-6.3) rectangle (0.15+2,0.45+3.3-6.3);	  

	  \draw [pattern=dots] (0.15+2,0+3.1) rectangle (0.15+2+2,0.45+3.1);
	  \draw [pattern=dots] (0.15+2,0+3.3-3) rectangle (0.15+2+2,0.45+3.3-3);
	  \draw [pattern=dots] (0.15+2,0+3.3-6.3) rectangle (0.15+2+2,0.45+3.3-6.3);	  
	  
	  \draw [pattern=north west lines] (0.15+2+2,0+3.1) rectangle (0.15+2+2+2,0.45+3.1);
	  \draw [pattern=north west lines] (0.15+2+2,0+3.3-3) rectangle (0.15+2+2+2,0.45+3.3-3);
	  \draw [pattern=north west lines] (0.15+2+2,0+3.3-6.3) rectangle (0.15+2+2+2,0.45+3.3-6.3);
	  
	  \draw [white,fill=black!15] (0.15+2+2+2,0+3.1) rectangle (0.15+2+2+2+2,0.45+3.1);
	  \draw [white,fill=black!15] (0.15+2+2+2,0+3.3-3) rectangle (0.15+2+2+2+2,0.45+3.3-3);
	  \draw [white,fill=black!15] (0.15+2+2+2,0+3.3-6.3) rectangle (0.15+2+2+2+2,0.45+3.3-6.3);	  	  

	  \draw [-] [dashed] (0.15,-0.8-3) -- (0.15,4.4);
	  \draw [-] [dashed] (0.15+2,-0.8-3) -- (0.15+2,4.4);
	  \draw [-] [dashed] (0.15+2+2,-0.8-3) -- (0.15+2+2,4.4);
	  \draw [-] [dashed] (0.15+2+2+2,-0.8-3) -- (0.15+2+2+2,4.4);	  
	  \draw [-] [dashed] (0.15+2+2+2+2,-0.8-3) -- (0.15+2+2+2+2,4.4);	  

	  \draw [-][very thick] (-0.1,0.44+3.1) -- (8.5,0.44+3.1);
	  \draw [-][very thick] (-0.1,0+3.1) -- (8.5,0+3.1);

	  \draw [-] [very thick] (0.15,0+3.1) -- (0.15,0.44+3.1);
	  \draw [-] [very thick] (0.15+2,0+3.1) -- (0.15+2,0.44+3.1);
	  \draw [-] [very thick] (0.15+2+2,0+3.1) -- (0.15+2+2,0.44+3.1);
	  \draw [-] [very thick] (0.15+2+2+2,0+3.1) -- (0.15+2+2+2,0.44+3.1);	  
	  \draw [-] [very thick] (0.15+2+2+2+2,0+3.1) -- (0.15+2+2+2+2,0.44+3.1);	  	  
	  
	  \draw [-] (0.15+2,0.44+3.1) -- (0.15+2,0.44+3.1+0.25);	  
	  \draw [-] (0.15+2+2,0.44+3.1) -- (0.15+2+2,0.44+3.1+0.25);	  
	  \draw [-] (0.15+2+2+2,0.44+3.1) -- (0.15+2+2+2,0.44+3.1+0.25);	  
	  \draw [-] (0.15+2+2+2+2,0.44+3.1) -- (0.15+2+2+2+2,0.44+3.1+0.25);	  

 	  \node [above] at (0.15+1,-0.4+3.5) {$\mathbf{v}_1$};

 	  \draw [white, fill=white] (6.9-4,-0.45+3.5+0.13) rectangle (6.9+0.5-4,-0.45+3.5+0.42);
	  \node [above] at (0.15+1+2,-0.4+3.5) {$\mathbf{v}_2$};
	  
	  \draw [white, fill=white] (4.2+0.65,-0.45+3.5+0.13) rectangle (4.2+0.5+0.65,-0.45+3.5+0.42);
	  \node [above] at (0.15+1+2+2,-0.4+3.5) {$\mathbf{v}_3$};	  
	  
	  \node [above] at (0.15+1+2+2+2,-0.4+3.5) {$\mathbf{v}_4$};	  	  

	  \draw [-] (0.15,0.55+3.1) -- (0.15+2,0.55+3.1);
	  \draw [myptr2] (0.15,0.55+3.1) -- (0.15+2,0.55+3.1);
	  \draw [myptr2] (0.15+2,0.55+3.1) -- (0.15,0.55+3.1);
	  
	  \draw [-] (0.15+2,0.55+3.1) -- (0.15+2+2,0.55+3.1);
	  \draw [myptr2] (0.15+2,0.55+3.1) -- (0.15+2+2,0.55+3.1);
	  \draw [myptr2] (0.15+2+2,0.55+3.1) -- (0.15+2,0.55+3.1);
	  
	  \draw [-] (0.15+2+2,0.55+3.1) -- (0.15+2+2+2,0.55+3.1);
	  \draw [myptr2] (0.15+2+2,0.55+3.1) -- (0.15+2+2+2,0.55+3.1);	  
	  \draw [myptr2] (0.15+2+2+2,0.55+3.1) -- (0.15+2+2,0.55+3.1);	  
	  
	  \draw [-] (0.15+2+2+2,0.55+3.1) -- (0.15+2+2+2+2,0.55+3.1);
	  \draw [myptr2] (0.15+2+2+2,0.55+3.1) -- (0.15+2+2+2+2,0.55+3.1);	  
	  \draw [myptr2] (0.15+2+2+2+2,0.55+3.1) -- (0.15+2+2+2,0.55+3.1);	  

	  \node [above] at (1,3.77) {$n_0$};
 	  \node [above] at (1+2,3.77) {$n_0$};
 	  \node [above] at (1+2+2,3.77) {$n_0$};
 	  \node [above] at (1+2+2+2,3.77) {$n_0$};

	  \draw [->,ultra thick] (0.15,3.1+0.44+0.6) -- (0.15,3.1+0.44);
 	  \draw [myptr1] (0.15,3.1+0.44+0.6) -- (0.15,3.1+0.44);
 	  \node [above] at (0.3,3.1+0.44+0.6) {$s = s_0$};

	  \draw [white, fill=white] (2.4,-0.9+3.3-0.25) rectangle (2.4+1.5+1.5,-0.9+3.3+0.25);
	  \node [right] at (2.4,-0.9+3.3) {(a) $n = n_0$, $s = s_0$};
	  
	  \draw [-][very thick] (-0.1,0.44+3.3-3) -- (8.5,0.44+3.3-3);
	  \draw [-][very thick] (-0.1,0+3.3-3) -- (8.5,0+3.3-3);

	  \draw [-] [very thick] (0.4,0+3.3-3) -- (0.4,0.44+3.3-3);	  
	  \draw [-] [very thick] (0.4+1.4,0+3.3-3) -- (0.4+1.4,0.44+3.3-3);	  	  
	  \draw [-] [very thick] (0.4+1.4+1.4,0+3.3-3) -- (0.4+1.4+1.4,0.44+3.3-3);	  	  
	  \draw [-] [very thick] (0.4+1.4+1.4+1.4,0+3.3-3) -- (0.4+1.4+1.4+1.4,0.44+3.3-3);	  	  	  
	  \draw [-] [very thick] (0.4+1.4+1.4+1.4+1.4,0+3.3-3) -- (0.4+1.4+1.4+1.4+1.4,0.44+3.3-3);	  	  
	  \draw [-] [very thick] (0.4+1.4+1.4+1.4+1.4+1.4,0+3.3-3) -- (0.4+1.4+1.4+1.4+1.4+1.4,0.44+3.3-3);	  	  
	  
	  \draw [-] (0.4+1.4,0.44+3.3-3) -- (0.4+1.4,0.44+3.3+0.25-3);	  	  
	  \draw [-] (0.15+2,0.44+3.3-3) -- (0.15+2,0.44+3.3+0.25-3);	  	  
	  \draw [-] (0.4+1.4+1.4,0.44+3.3-3) -- (0.4+1.4+1.4,0.44+3.3+0.25-3);	  	  
	  \draw [-] (0.4+1.4+1.4+1.4,0.44+3.3-3) -- (0.4+1.4+1.4+1.4,0.44+3.3+0.25-3);	  	  
	  \draw [-] (0.4+1.4+1.4+1.4+1.4,0.44+3.3-3) -- (0.4+1.4+1.4+1.4+1.4,0.44+3.3+0.25-3);	  	  
	  \draw [-] (0.4+1.4+1.4+1.4+1.4+1.4,0.44+3.3-3) -- (0.4+1.4+1.4+1.4+1.4+1.4,0.44+3.3+0.25-3);	  	  
	  
	  \draw [-] (0.4,0.55+3.3-3) -- (0.4+1.4,0.55+3.3-3);	  
	  \draw [myptr2] (0.4,0.55+3.3-3) -- (0.4+1.4,0.55+3.3-3);	  
	  \draw [myptr2] (0.4+1.4,0.55+3.3-3) -- (0.4,0.55+3.3-3);	  
	  
	  \draw [-] (0.15+2,0.55+3.3-3) -- (0.4+1.4,0.55+3.3-3);
	  \draw [myptr2] (0.15+2,0.55+3.3-3) -- (0.4+1.4,0.55+3.3-3);	  	  
	  \draw [myptr2] (0.4+1.4,0.55+3.3-3) -- (0.15+2,0.55+3.3-3);	  	  
	  
	  \draw [-] (0.15+2,0.55+3.3-3) -- (0.4+1.4+1.4,0.55+3.3-3);
	  \draw [myptr2] (0.15+2,0.55+3.3-3) -- (0.4+1.4+1.4,0.55+3.3-3);
	  \draw [myptr2] (0.4+1.4+1.4,0.55+3.3-3) -- (0.15+2,0.55+3.3-3);
	  
	  \draw [-] (0.4+1.4+1.4,0.55+3.3-3) -- (0.4+1.4+1.4+1.4,0.55+3.3-3);
	  \draw [myptr2] (0.4+1.4+1.4,0.55+3.3-3) -- (0.4+1.4+1.4+1.4,0.55+3.3-3);
	  \draw [myptr2] (0.4+1.4+1.4+1.4,0.55+3.3-3) -- (0.4+1.4+1.4,0.55+3.3-3);
	  
	  \draw [-] (0.4+1.4+1.4+1.4,0.55+3.3-3) -- (0.4+1.4+1.4+1.4+1.4,0.55+3.3-3);
	  \draw [myptr2] (0.4+1.4+1.4+1.4,0.55+3.3-3) -- (0.4+1.4+1.4+1.4+1.4,0.55+3.3-3);	  
	  \draw [myptr2] (0.4+1.4+1.4+1.4+1.4,0.55+3.3-3) -- (0.4+1.4+1.4+1.4,0.55+3.3-3);	  
	  
	  \draw [-] (0.4+1.4+1.4+1.4+1.4,0.55+3.3-3) -- (0.4+1.4+1.4+1.4+1.4+1.4,0.55+3.3-3);
	  \draw [myptr2] (0.4+1.4+1.4+1.4+1.4,0.55+3.3-3) -- (0.4+1.4+1.4+1.4+1.4+1.4,0.55+3.3-3);	  
	  \draw [myptr2] (0.4+1.4+1.4+1.4+1.4+1.4,0.55+3.3-3) -- (0.4+1.4+1.4+1.4+1.4,0.55+3.3-3);	  
	  
 	  \node [above] at (0.4+0.7,0.44+3.3-3+0.15) {$n$}; 	  	  
 	  \node [above] at (0.4+1.3+0.3,0.44+3.3-3+0.15) {$d_1^{\prime}$}; 	  	   	  
 	  \node [above] at (0.4+0.8+1.4,0.44+3.3-3+0.1) {$d_2^{\prime}$}; 	  	   	   	  
 	  \node [above] at (0.4+0.7+1.4+1.4,0.44+3.3-3+0.15) {$n$};  	  
 	  \node [above] at (0.4+0.7+1.4+1.4+1.4,0.44+3.3-3+0.15) {$n$};  	   	  
 	  \node [above] at (0.4+0.7+1.4+1.4+1.4+1.4,0.44+3.3-3+0.15) {$n$};  	  
 	  
 	  \draw[decorate,decoration=brace] (0.4+1.4,0+3.3-3-0.15)--(0.4,0+3.3-3-0.15);
 	  \draw[decorate,decoration=brace] (0.4+1.4+1.4,0+3.3-3-0.15)--(0.4+1.4,0+3.3-3-0.15);
 	  \draw[decorate,decoration=brace] (0.4+1.4+1.4+1.4,0+3.3-3-0.15)--(0.4+1.4+1.4,0+3.3-3-0.15);
 	  \draw[decorate,decoration=brace] (0.4+1.4+1.4+1.4+1.4,0+3.3-3-0.15)--(0.4+1.4+1.4+1.4,0+3.3-3-0.15);
 	  \draw[decorate,decoration=brace] (0.4+1.4+1.4+1.4+1.4+1.4,0+3.3-3-0.15)--(0.4+1.4+1.4+1.4+1.4,0+3.3-3-0.15);

 	  \node [above] at (0.4+0.7,0.44+3.3-3+0.15-1.4) {$\mathbf{w}_1$};  	  
 	  \node [above] at (0.4+0.7+1.4,0.44+3.3-3+0.15-1.4) {$\mathbf{w}_2$};  	  
 	  \node [above] at (0.4+0.7+1.4+1.4,0.44+3.3-3+0.15-1.4) {$\mathbf{w}_3$};  	  
 	  \node [above] at (0.4+0.7+1.4+1.4+1.4,0.44+3.3-3+0.15-1.4) {$\mathbf{w}_4$};  	  
 	  \node [above] at (0.4+0.7+1.4+1.4+1.4+1.4,0.44+3.3-3+0.15-1.4) {$\mathbf{w}_5$};  	  
  
	  \draw [->,ultra thick] (0.4,3.3+0.44+0.6-3) -- (0.4,3.3+0.44-3);
 	  \node [above] at (0.4,3.3+0.44+0.6-3) {$s \neq s_0$};  
	  
	  \draw [white, fill=white] (2.4+1.5,-1.3+3.3-3.1-0.25) rectangle (2.4+1.5+3,-1.3+3.3-3.1+0.25);
	  \node [right] at (2.4,-1.3+3.3-3.1) {(b) $n < n_0$, $s \neq s_0$};
	  
	  \draw [-][very thick] (-0.1,0.44+3.3-6.3) -- (8.5,0.44+3.3-6.3);
	  \draw [-][very thick] (-0.1,0+3.3-6.3) -- (8.5,0+3.3-6.3);

	  \draw [-] [very thick] (1.3,0+3.3-6.3) -- (1.3,0.44+3.3-6.3);	  
	  \draw [-] [very thick] (1.3+3.3,0+3.3-6.3) -- (1.3+3.3,0.44+3.3-6.3);	  	  
	  \draw [-] [very thick] (1.3+3.3+3.3,0+3.3-6.3) -- (1.3+3.3+3.3,0.44+3.3-6.3);	  	

	  \draw [-] (0.15+2,0.44+3.3-6.3) -- (0.15+2,0.44+3.3-6.3+0.25);	  
	  \draw [-] (1.3+3.3,0.44+3.3-6.3) -- (1.3+3.3,0.44+3.3-6.3+0.25);	  
	  \draw [-] (1.3+3.3+3.3,0.44+3.3-6.3) -- (1.3+3.3+3.3,0.44+3.3-6.3+0.25);	  
	  
	  \draw [myptr2] (1.3,0.55+3.3-6.3) -- (0.15+2,0.55+3.3-6.3);
	  \draw [myptr2] (0.15+2,0.55+3.3-6.3) -- (1.3,0.55+3.3-6.3);
	  
	  \draw [-] (1.3+3.3,0.55+3.3-6.3) -- (0.15+2,0.55+3.3-6.3);
	  \draw [myptr2] (1.5+3.3,0.55+3.3-6.3) -- (0.15+2,0.55+3.3-6.3);
	  \draw [myptr2] (0.15+2,0.55+3.3-6.3) -- (0.15+2+2,0.55+3.3-6.3);
	  
	  \draw [-] (1.3+3.3,0.55+3.3-6.3) -- (0.15+2+2,0.55+3.3-6.3);
	  \draw [myptr2] (1.3+3.3,0.55+3.3-6.3) -- (0.15+2+2,0.55+3.3-6.3);	  
	  \draw [myptr2] (0.15+2+2,0.55+3.3-6.3) -- (1.3+3.3,0.55+3.3-6.3);	  
	  
	  \draw [-] (1.3+3.3,0.55+3.3-6.3) -- (1.3+3.3+3.3,0.55+3.3-6.3);
	  \draw [myptr2] (1.3+3.3,0.55+3.3-6.3) -- (1.3+3.3+3.3,0.55+3.3-6.3);
	  \draw [myptr2] (1.3+3.3+3.3,0.55+3.3-6.3) -- (1.3+3.3,0.55+3.3-6.3);
	  
 	  \node [above] at (1+0.7,0.44+3.3-6.3+0.15) {$d_1$}; 	  
 	  \node [above] at (1+2.1,0.44+3.3-6.3+0.1) {$n_0$}; 	   	  
 	  \node [above] at (1+0.4+1+2,0.44+3.3-6.3+0.1) {$d_2$}; 	   	   	  
 	  \node [above] at (1.3+3.3+1.65,0.44+3.3-6.3+0.15) {$n$}; 	   	   	  

 	  \draw[decorate,decoration=brace] (1.3+3.3,0+3.3-6.3-0.15)--(1.3,0+3.3-6.3-0.15); 	  
 	  \draw[decorate,decoration=brace] (1.3+3.3+3.3,0+3.3-6.3-0.15)--(1.3+3.3,0+3.3-6.3-0.15); 	   	  
 	  
 	  \node [above] at (1.3+1.65,0.44+3.3-6.3+0.15-1.4) {$\mathbf{w}_1$};  	   	  
 	  \node [above] at (1.3+1.65+3.3,0.44+3.3-6.3+0.15-1.4) {$\mathbf{w}_2$};  	   	  
 	  
	  \draw [->,ultra thick] (1.3,3.3+0.44+0.6-6.3) -- (1.3,3.3+0.44-6.3);
 	  \node [above] at (1.3,3.3+0.44+0.6-6.3) {$s \neq s_0$};  
	  
	  \node [right] at (2.2,-1.3+3.3-6.5) {(c) $n > n_0$, $n \neq ln_0$, and $s \neq s_0$};	  

    \end{tikzpicture}
\end{center}

\caption{A binary cyclic code $C(n_0,g_0)$ is used at the transmitter and 
$\mathbf{v}_1, \mathbf{v}_2, \ldots, \mathbf{v}_4 \in C(n_0,g_0)$.
Figures~(a), (b), and (c) correspond to the situations when $n=n_0$, $s = s_0$, $n<n_0$, $s \neq s_0$, and
$n>n_0$, $s \neq s_0$ respectively.}
\label{Figure_n_more_less_n0_noise_free}
\end{figure}
%

We first consider the case when either $n \neq ln_0$ or $s \neq s_0$, where $l \in \mathbb{N}$ .
The case when $n = ln_0$ and $s = s_0$ will be studied towards the end of this section.
Consider a noise-free sequence of codewords of the true code $C(n_0,g_0)$ as shown in \Fig~\ref{Figure_n_more_less_n0_noise_free}(a).
Example situations when this sequence is divided into vectors of length $n$ such that $n < n_0$, $s \neq s_0$ and $n > n_0$, $s \neq s_0$
are illustrated in Figures~\ref{Figure_n_more_less_n0_noise_free}(b) and (c) respectively.
In \Fig~\ref{Figure_n_more_less_n0_noise_free}(a), $\mathbf{v}_1, \mathbf{v}_2, \ldots, \mathbf{v}_4 \in C(n_0,g_0)$.
In \Fig~\ref{Figure_n_more_less_n0_noise_free}(b), $\mathbf{w}_1, \mathbf{w}_2, \ldots, \mathbf{w}_5$
are vectors of length $n < n_0$ and 
in \Fig~\ref{Figure_n_more_less_n0_noise_free}(c), $\mathbf{w}_1$ and $\mathbf{w}_2$
are vectors of length $n > n_0$.
In this section, we use the alphabets $\mathbf{v}$ and $\mathbf{w}$ to denote the vectors of lengths $n_0$ 
and $n$ respectively.

From \Fig~\ref{Figure_n_more_less_n0_noise_free}(b) and (c),
it can be seen that when either $n \neq ln_0$ or $s \neq s_0$, a vector $\mathbf{w}_j$ of length $n$ is either of the following two types.
\begin{enumerate}
\item Vector $\mathbf{w}_j$ is formed by the consecutive $n$ bits of some codeword in code $C(n_0,g_0)$.
For example, vectors $\mathbf{w}_1$ and $\mathbf{w}_4$ in \Fig~\ref{Figure_n_more_less_n0_noise_free}(b)
are formed by the consecutive $n$ bits of codewords $\mathbf{v}_1$ and $\mathbf{v}_3$ of $C(n_0,g_0)$ respectively.
\item Vector $\mathbf{w}_j$ is formed by the concatenation of the suffix, a sequence of $q$ codewords, and the prefix of
a codeword in the true code, where $q \in \mathbb{Z}, q \geq 0$.
For example, $\mathbf{w}_1$ in \Fig~\ref{Figure_n_more_less_n0_noise_free}(c) is formed by the concatenation of
the suffix of $\mathbf{v}_1$ of length $d_1$, $\mathbf{v}_2$, and the prefix of $\mathbf{v}_3$ of length $d_2$, where $0 \leq d_1,d_2 < n_0$
such that $n = d_1+n_0+d_2$.
The vector $\mathbf{w}_2$ in \Fig~\ref{Figure_n_more_less_n0_noise_free}(b) is formed by the concatenation of
the suffix of $\mathbf{v}_2$ of length $d_1^{\prime}$ and the prefix of $\mathbf{v}_3$ of length $d_2^{\prime}$, 
such that $n=d_1^{\prime}+d_2^{\prime}$.
\end{enumerate}

We denote the vector $\mathbf{w}_j$ of the second type by $\mathbf{w}_j^{\prime}$ to distinguish between the $n$ bits vectors
of the two types mentioned above.
For the simplicity of notation, we will ignore suffix $j$ from $\mathbf{w}_j$ and $\mathbf{w}_j^{\prime}$.
Using this notation, $\mathbf{w}$ is an $n$-bit vector formed by the consecutive $n$ bits of a codeword in $C(n_0,g_0)$.
Since $C(n_0,g_0)$ is a cyclic code, it is sufficient to consider the case when $\mathbf{w}$ is formed by 
the initial $n$ bits of a codeword in $C(n_0,g_0)$, i.e., $\mathbf{w}$ is given by,
\begin{align}
\mathbf{w} = \mathbf{v}(0:n-1),
\label{Eqn_w_definition}
\end{align}
where $\mathbf{v} \in C(n_0,g_0)$.
Let $\mathcal{W}(n)$ be the linear subspace obtained by puncturing the last $n_0-n$ bits of codewords of code $C(n_0,g_0)$.
It follows that $\mathbf{w} \in \mathcal{W}(n)$. 

As explained in the previous paragraph, $\mathbf{w}^{\prime}$ is an $n$-bit vector formed by 
the concatenation of the suffix of length $d_1$, a sequence of $q$ codewords, and the prefix of
length $d_2$, where $d_1,d_2, q \in \mathbb{N}$, 
such that $n = d_1+qn_0+d_2$, i.e., $\mathbf{w}^{\prime}$ is given by,
\begin{align}
\mathbf{w}^{\prime} &= \Big[ \mathbf{v}_{1}(n_0-d_1:n_0-1)  \mbox{~} \mathbf{v}_{2} \mbox{~} \cdots \mbox{~} \mathbf{v}_{q+1}  \mbox{~} \mathbf{v}_{q+2}(0:d_2-1) \Big]  
\label{Eqn_w_prime_definition}
\end{align}
where $\mathbf{v}_1, \mathbf{v}_2, \ldots, \mathbf{v}_{q+2} \in C(n_0,g_0)$.
Let $C_1(d_1)$ and $C_2(d_2)$ be the linear block codes obtained by 
considering the set of suffixes and prefixes of lengths $d_1$ and $d_2$ of codewords in $C(n_0,g_0)$ respectively.
Let $\mathcal{W}^{\prime}(n)$ be the linear subspace obtained by concatenating 
all possible suffixes of length $d_1$, $q$ codewords, and prefixes of length $d_2$, i.e., 
\begin{align}
\mathcal{W}^{\prime}(n) & \coloneqq  C_1(d_1) + 
\underbrace{ C(n_0,g_0) + \mbox{~} \cdots \mbox{~} + C(n_0,g_0)}_{q \text{ times}} + C_2(d_2).
\label{Eqn_Wn_prime_pre_suff_definition}
\end{align}
From (\ref{Eqn_w_prime_definition}), it can be seen that $\mathbf{w}^{\prime} \in \mathcal{W}^{\prime}(n)$.
Note that, since every codeword in $C(n_0,g_0)$ is chosen according to the uniform
distribution, any $\mathbf{w} \in \mathcal{W}(n)$ and $\mathbf{w}^{\prime} \in \mathcal{W}^{\prime}(n)$ 
occur with the uniform distribution over the set of codewords in $\mathcal{W}(n)$ and $\mathcal{W}^{\prime}(n)$ respectively.

For a factor $f(X)$ of $X^n+1$, suppose $r(X)=\mathbf{w}(X) \bmod f(X)$ and $r^{\prime}(X)=\mathbf{w}^{\prime}(X) \bmod f(X)$.
From \Fig~\ref{Figure_n_more_less_n0_noise_free}, in order to study the syndrome distribution of the noise-free sequence,
we need to study the distributions $r(X)$ and $r^{\prime}(X)$.
In the following proposition, we first prove that the distributions of $r(X)$ and $r^{\prime}(X)$ can 
either be uniform or restricted uniform.

%
\begin{proposition}
\label{Proposition_cyclic_structure_no_type1}
For a cyclic code $C(n_0,g_0)$, let $\mathcal{W}(n)$ and $\mathcal{W}^{\prime}(n)$ be the linear subspaces
as defined in the previous paragraph, where $n$ is not a multiple of $n_0$.
For a factor $f(X)$ of $X^n+1$, suppose $r(X) = \mathbf{w}(X) \bmod f(X)$
and $r^{\prime}(X) = \mathbf{w}^{\prime}(X) \bmod f(X)$, where $\mathbf{w}(X) \in \mathcal{W}(n)$ 
and $\mathbf{w}^{\prime}(X) \in \mathcal{W}^{\prime}(n)$.
Then the random variables corresponding to $r(X)$ and $r^{\prime}(X)$ can either follow the uniform distribution or the restricted uniform distribution.
(see (\ref{Eqn_type_2_distri}), (\ref{Eqn_type_3_distri}), \Fig~\ref{Figure_example_distributions}).
\end{proposition}
\begin{IEEEproof}
The proof is given in Appendix~B.
\end{IEEEproof}

Note that Proposition~\ref{Proposition_cyclic_structure_no_type1} is true irrespective of whether $f(X)$ is a factor of $g_0(X)$ or not.
This proposition says that the distribution of $r(X)$ and $r^{\prime}(X)$
can be either be uniform or restricted uniform, but it does not specify when the distribution will be of either of the type.
In the next two sections we will answer this question.

\subsection{Analyzing the distribution of $r(X)$}
\label{SubSection_cyclic_strure_w_consecutive_n_bits}

In this section, we characterize the distribution of $r(X) = \mathbf{w}(X) \bmod f(X)$, when
$\mathbf{w}(X)$ is formed by the $n$ consecutive bits of a codeword in $C(n_0,g_0)$.
Due to the cyclic nature of the code, it is sufficient to consider the case when $\mathbf{w}(X)$ is formed by 
the initial $n$ bits of a codeword in $C(n_0,g_0)$.
Depending on the chosen $n$ and the degree of $f(X)$ we have the following cases.
\begin{enumerate}[(a)]
\item $n \leq k_0$, where recall that $k_0$ is the dimension of $C(n_0,k_0)$\\
When $n \leq k_0$, a vector $\mathbf{w}$ formed the initial $n$ bits of a codeword in $C(n_0,g_0)$
can take all possible $2^n$ values in $\mathbb{F}_2^n$ 
since, for a cyclic code any set of $k_0$ consecutive coordinate locations form an information set~\cite{Huffman_Pless_ECC}.
From our system assumption, any codeword in $C(n_0,g_0)$ is chosen i.i.d.~according to the uniform
distribution.
Hence $\mathbf{w}(X)$ will take all possible values in $\mathcal{P}_n$ with equal probability and
the random variable corresponding to $r(X)$ will follow the uniform distribution.
\item $\deg(f) > k_0$\\
The syndrome $r(X) = \mathbf{w}(X) \bmod f(X)$ can take $2^{\deg(f)}$ possible values in $\mathcal{P}_{\deg(f)}$. 
Whereas, $\mathbf{w}(X)$ can take at most $2^{k_0}$ possible values.
When $\deg(f) > k_0$, the number of possible syndromes are more than the number of possible $\mathbf{w}(X)$.
This implies that the random variable corresponding to $r(X)$ cannot follow the uniform distribution
and from Proposition~\ref{Proposition_cyclic_structure_no_type1}, $r(X)$ follows the restricted uniform distribution. 
\item $k_0 < n < n_0$ and $\deg(f) \leq k_0$\\
In this case, the distribution of $r(X)$ can either be uniform or restricted uniform.
We characterize the conditions under which the restricted uniform distribution is possible in the following theorem.
\end{enumerate}
%

%
\begin{theorem}
\label{Theorem_main_iff_structure}
Consider a non-degenerate cyclic code $C(n_0,g_0)$ of length $n_0$, dimension $k_0$, and generator polynomial $g_0(X)$.
Let $g_0^{\perp}(X)$ be the generator polynomial of the dual code of $C(n_0,g_0)$.
For an integer $n$ and $\mathbf{v}(X) \in C(n_0,g_0)$, suppose $\mathbf{w}(X) = \mathbf{v}(X) \bmod X^n$ such that $k_0 < n < n_0$.
Suppose $f(X)$ is a factor of $X^n+1$ such that $\deg(f) \leq k_0$ and $r(X) = \mathbf{w}(X) \bmod f(X)$.
Then the necessary condition for the random variable corresponding to $r(X)$ to follow the restricted uniform distribution is 
that $g_0^{\perp}(X)$ should have a factor of order strictly less than $n_0$.

Conversely, when $g_0^{\perp}(X)$ has a factor $m^{\perp}(X)$ of order $n^{\prime}$ such that $1 \leq n^{\prime} < n_0$, 
syndrome 
$r(X)$ follows the restricted uniform distribution if the chosen $n$ and $f(X)$ satisfy the following conditions.
\begin{enumerate}
\item $n=bn^{\prime}$ for some $b \in \mathbb{N}$.
\item $f(X)$ is a factor of $m(X)(1+X^{n^{\prime}}+X^{2n^{\prime}}+\ldots+X^{(b-1)n^{\prime}})$, where 
$m(X)$ is the minimal generating polynomial of the linear recurring sequence whose minimal polynomial is $m^{\perp}(X)$ 
such that $\deg(m^{\perp}) > k_0-\deg(f)$ (see Definition~\ref{Definition_min_gen_poly_LRR}).
\end{enumerate}
\end{theorem}
\begin{IEEEproof}
The proof is given in Appendix~C. 
\end{IEEEproof}

We next provide an example of a cyclic code that satisfies the claim of this theorem.
\begin{example}
\label{Example_thm_iff_main}
Consider a non-degenerate cyclic code $C(15,g_0)$ with generator polynomial $g_0(X) = (X^4+X^3+1)(X^4+X^3+X^2+X+1)(X+1)$
and dimension $k_0 = 6$. 
The generator polynomial of the dual code of $C(15,g_0)$ is 
$g_0^{\perp}(X) = (X^2 + X + 1) (X^4 + X^3 + 1)$. Note that the factor $m^{\perp}(X) = X^2 + X + 1$
of $g_0^{\perp}(X)$ has the order $n^{\prime}=3$, which is strictly less than $n_0=15$. 
The minimal generating polynomial corresponding to $m^{\perp}(X) = X^2 + X + 1$ is $m(X) = X+1$.

For $n=9$ and $f(X) = X^6+X^3+1$, by considering all possible codewords in $C(15,g_0)$
it can be checked that the random variable corresponding to $r(X) = \mathbf{w}(X) \bmod f(X)$ follows 
the restricted uniform distribution\footnote{For this $n$ and $f$, the probability of zero syndrome is $0.0625$. 
For the uniform distribution,
the probability of zero syndrome would be $1/2^{\deg(f)} = 1/2^6 = 0.015625$.}.
Note that the chosen $n$ and $f(X)$ satisfy the conditions of the theorem as
$n=3n^{\prime}$, i.e., $b=3$ and $f(X) = X^6+X^3+1$ is a factor 
of $m(X)(1+X^{n^{\prime}}+X^{2n^{\prime}}+\ldots+X^{(b-1)n^{\prime}}) = (X+1) (1+X^3+X^6)$
such that $\deg(m^{\perp}) > k_0-\deg(f)$.
\hfill $\square$
\end{example}
%

\subsection{Analyzing the distribution of $r^{\prime}(X)$}
\label{SubSection_cyclic_strure_w_prefix_suffix}

In this section, we study the distribution of $r^{\prime}(X) = \mathbf{w}^{\prime}(X) \bmod f(X)$, 
where $\mathbf{w}^{\prime} \in \mathcal{W}^{\prime}(n)$ (see (\ref{Eqn_w_prime_definition}) and (\ref{Eqn_Wn_prime_pre_suff_definition})).
We first consider the case when $n < n_0$ and $r(X)$ follows the uniform distribution.
In the following proposition, we will prove that when $r(X)$ follows the uniform distribution, 
$r^{\prime}(X)$ also follows the uniform distribution.

\begin{theorem}
\label{Theorem_W_suff_pre_cyclic_structure}
Suppose assumed length $n$ is strictly less than the true length $n_0$ of the code.
Let $r(X) = \mathbf{w}(X) \bmod f(X)$ and $r^{\prime}(X) = \mathbf{w}^{\prime}(X) \bmod f(X)$,
where $\mathbf{w}$ and $\mathbf{w}^{\prime}$ are defined in (\ref{Eqn_w_definition})
and (\ref{Eqn_w_prime_definition}) respectively.
Then the random variable corresponding to $r^{\prime}(X)$ follows the uniform distribution if
the random variable corresponding to $r(X)$ follows the uniform distribution.
\end{theorem}
\begin{IEEEproof}
The proof is given in Appendix~D. 
\end{IEEEproof}
%
We next consider the case when either $n > n_0$ or $r(X)$ follows the restricted uniform distribution.
From Proposition~\ref{Proposition_cyclic_structure_no_type1}, we know that $r^{\prime}(X)$ will follow the 
uniform distribution or restricted uniform distribution.
We now provide the conditions under which $r^{\prime}(X)$ will follow the uniform and 
the restricted uniform distributions.
From (\ref{Eqn_w_prime_definition}), $\mathbf{w}^{\prime}$ is given by,
\begin{equation}
\begin{aligned}
\mathbf{w}^{\prime} &=\Big[ \mathbf{v}_{1}(n_0-d_1:n_0-1)  \mbox{~} \mathbf{v}_{2} \mbox{~} \cdots \mbox{~} \mathbf{v}_{q+1}  \mbox{~} \mathbf{v}_{q+2}(0:d_2-1) \Big] \\ 
		    & = \Big[ \mathbf{c}_{1}  \mbox{~~} \mathbf{v}_{2} \mbox{~} \cdots \mbox{~} \mathbf{v}_{q+1}  \mbox{~~}  \mathbf{c}_{2} \Big], 
\label{Eqn_w_prime_def23}
\end{aligned}
\end{equation}
where $\mathbf{v}_i \in C(n_0,g_0)$ for $i = 1,2,\ldots, q+2$, 
$\mathbf{c}_{1} \coloneqq \mathbf{v}_{1}(n_0-d_1:n_0-1)$, and $\mathbf{c}_{2} \coloneqq \mathbf{v}_{q+2}(0:d_2-1)$.
From (\ref{Eqn_w_prime_def23}), $r^{\prime}(X)$ is given by,
\begin{equation}
\begin{aligned}
r^{\prime}(X) &= \mathbf{w}^{\prime}(X) \bmod f(X) \\
	      &= \Big[ \mathbf{c}_1(X) + X^{d_1} \mathbf{v}_2(X) + \ldots + X^{d_1+(q-1)n_0}\mathbf{v}_{q+1}(X) \\
	      & \mbox{~~~~~~~~~~~~~~~~~~~~~~~~~~~}+ X^{d_1+qn_0}\mathbf{c}_{2}(X) \Big] \bmod f(X) \\
	      &= t_1(X) + t_2(X) + \ldots + t_{q+2}(X), 
\end{aligned}
\label{Eqn_t1_t2_tq2}
\end{equation}
where $t_1(X) = \mathbf{c}_1(X) \bmod f(X)$, $t_i(X) = X^{d_1+(i-2)n_0} \mathbf{v}_{i}(X) \bmod f(X)$, and
$t_{q+2}(X) = X^{d_1+qn_0} \mathbf{c}_2(X) \bmod f(X)$, for $i = 2,3,\ldots, q+1$.
The distribution of $t_1(X)$ and $t_{q+2}(X)$ can be studied using Section~\ref{SubSection_cyclic_strure_w_consecutive_n_bits},
since $\mathbf{c}_1(X)$ and $\mathbf{c}_2(X)$ are formed by the consecutive $d_1$ and $d_2$ bits of a codeword in $C(n_0,g_0)$.
We now study the distribution of $t_i(X)$, for $i = 2,3,\ldots, q+1$.
First note that when $\mathbf{v}(X) \bmod f(X)$ follows the uniform distribution, 
$X^d \mathbf{v}(X) \bmod f(X)$ also follows the uniform distribution for any positive integer $d$.
Similarly, when $\mathbf{v}(X) \bmod f(X)$ follows the restricted uniform distribution, 
$X^d \mathbf{v}(X) \bmod f(X)$ also follows the restricted uniform distribution.
Hence it is sufficient to study the distribution of $\mathbf{v}(X) \bmod f(X)$.
\begin{itemize}
\item When $f(X)$ is a factor of $g_0(X)$, $\mathbf{v}(X) \bmod f(X)$ is zero with probability one,
since $\mathbf{v}(X) = \mathbf{u}(X) g_0(X)$ for some $\mathbf{u}(X) \in \mathcal{P}_{k_0}$.
Thus $\mathbf{v}(X) \bmod f(X)$ follows the degenerate distribution.
\item When $f(X)$ is not a factor of $g_0(X)$ and $\deg(f) \leq k_0$, from Theorem~3.2 of \cite{EuropeanWireless2014},
$\mathbf{v}(X) \bmod f(X)$ follows the uniform distribution.
\item When $f(X)$ is not a factor of $g_0(X)$ and $\deg(f) > k_0$, $\mathbf{v}(X) \bmod f(X)$ follows the restricted uniform distribution 
since the number of possible values that $r(X)$ can take are more than the number of possible values $\mathbf{v}(X)$
can take, as explained in Section~\ref{SubSection_cyclic_strure_w_consecutive_n_bits}.
\end{itemize}
We now study the distribution of $r^{\prime}(X)$ in the following theorem.

\begin{theorem}
\label{Theorem_structure_w_prime_type3}
Let $r^{\prime}(X)$ be as defined in (\ref{Eqn_t1_t2_tq2}).
Then $r^{\prime}(X)$ follows the uniform distribution when every $t_i(X)$, for $i = 1,2,\ldots,q+2$
follows the uniform distribution, otherwise it follows the restricted uniform distribution.
\end{theorem}
\begin{IEEEproof}
The proof is given in Appendix~E. 
\end{IEEEproof}
%

Theorems~\ref{Theorem_main_iff_structure}, \ref{Theorem_W_suff_pre_cyclic_structure}, and \ref{Theorem_structure_w_prime_type3} 
completely characterize the distribution of syndromes of the noise-free sequence when either $n \neq ln_0$ or $s \neq s_0$.
We next consider the case when $n=ln_0$ and $s=s_0$.

\subsection{The case when $n=ln_0$ and $s=s_0$}
\label{subsection_n_ln0_s_s0}

When $n=ln_0$ and $s=s_0$, every noise-free $n$-bit vector $\mathbf{w}_j$ is formed by the concatenation of $l$
codewords of the true code $C(n_0,g_0)$, i.e., any $\mathbf{w}_j$ for $1 \leq j \leq M$ is given by,
\begin{align}
\mathbf{w}_j = \big[ \mathbf{v}_1 \mbox{~~} \mathbf{v}_2 \mbox{~~} \ldots \mbox{~~} \mathbf{v}_l \big],
\label{Eqn_w_concat_v_l}
\end{align}
where $\mathbf{v}_i \in C(n_0,g_0)$ for $i = 1,2,\ldots,l$.
Depending on whether $f(X)$ is a factor of $g_0(X)$ or not and the degree of $f(X)$, we have the following cases.
\begin{enumerate}[(a)]
\item When $f(X)$ is a factor of $g_0(X)$, from (\ref{Eqn_w_concat_v_l}) $\mathbf{w}_j(X) \bmod f(X)$ is always zero since 
every $\mathbf{v}_i(X)$, for $i = 1,2,\ldots,l$ is a multiple of $g_0(X)$. This implies that $\mathbf{w}_j(X) \bmod f(X)$ follows the degenerate 
distribution.
\item When $f(X)$ is not a factor of $g_0(X)$ and $\deg(f) \leq k_0$, from Theorem~3.2 of \cite{EuropeanWireless2014},
$\mathbf{v}_i(X) \bmod f(X)$ follows the uniform distribution, for $i = 1,2,\ldots,l$. 
From Theorem~\ref{Theorem_structure_w_prime_type3} this implies that $\mathbf{w}_j(X) \bmod f(X)$
also follows the uniform distribution. 
\item When $f(X)$ is not a factor of $g_0(X)$ and $\deg(f) > k_0$, as explained in the previous section, each 
$\mathbf{v}_i(X) \bmod f(X)$ follows the restricted uniform distribution and
from Theorem~\ref{Theorem_structure_w_prime_type3}, $\mathbf{w}_j(X) \bmod f(X)$ follows the restricted uniform distribution. 
\end{enumerate}
%


\subsection{Summary of the distribution of $\mathbf{w}_j(X) \bmod f(X)$}
\label{subsection_sumary_distri_wj}

In this section, we summarize the results for the distribution of $\mathbf{w}_j(X) \bmod f(X)$.
Depending upon the chosen $n$, $s$, and $f(X)$, we have the following cases.
\begin{itemize}
\item When $n = ln_0$ for some $l \in \mathbb{N}$, $s=s_0$, and $f(X)$ is factor of $g_0(X)$, 
$\mathbf{w}_j(X) \bmod f(X)$ follows the degenerate distribution (see Section~\ref{subsection_n_ln0_s_s0} (a)).
\item When either $n \neq ln_0$ or $s \neq s_0$ or $f(X)$ is not a factor of $g_0(X)$, the distribution of $\mathbf{w}_j(X) \bmod f(X)$ is 
either uniform or restricted uniform. Theorems~\ref{Theorem_main_iff_structure}, \ref{Theorem_W_suff_pre_cyclic_structure}, 
and \ref{Theorem_structure_w_prime_type3} and Section~\ref{subsection_n_ln0_s_s0} (b), (c)
provide the conditions when the distribution is uniform or restricted uniform. 
\end{itemize}


\section{Syndrome distribution of the noise-affected received sequence}
\label{Section_cyclic_structure_noise_affected}
In the previous section, we studied the distribution of $\mathbf{w}_j(X) \bmod f(X)$, where $1 \leq j \leq M$.
In this section, we study the distribution of $\mathbf{y}_j(X) \bmod f(X)$, where recall that 
$\mathbf{y}_j(X)$ is the noise-affected version of $\mathbf{w}_j(X)$ (see (\ref{Equation_yj_wj_ej_temp})).
In the previous section, we proved that the distribution of $\mathbf{w}_j(X) \bmod f(X)$ is either degenerate or
uniform or restricted uniform.
%
We consider the case when $\mathbf{w}_j(X) \bmod f(X)$ follows each type of the distribution separately
and study the distribution of $\mathbf{y}_j(X) \bmod f(X)$.
The case when $\mathbf{w}_j(X) \bmod f(X)$ follows the degenerate distribution, i.e., when $n=ln_0$, $s=s_0$, and
$f(X)$ is a factor of $g_0(X)$ is studied in detail in \cite{EuropeanWireless2014} and \cite{TCOMM_2016}.
In the following theorem, we consider that case when $\mathbf{w}_j(X) \bmod f(X)$ follows the uniform distribution.

\begin{theorem}
\label{Theorem_wj_yj_equally_likely}
Let $\mathbf{y}_j(X)$ and $\mathbf{w}_j(X)$ be the $j$th noise-affected received polynomial and error-free polynomial 
respectively, for $j = 1,2,\ldots,M$.
Then $\mathbf{y}_j(X) \bmod f(X)$ follows the uniform distribution if $\mathbf{w}_j(X) \bmod f(X)$ follows the uniform distribution.
\end{theorem}
\begin{IEEEproof}
Since $\mathbf{w}_j(X) \bmod f(X)$ follows the uniform distribution, it takes any value in $\mathcal{P}_{\deg(f)}$
with probability $1/2^{\deg(f)}$. We now find the probability that $\mathbf{y}_j(X) \bmod f(X)$ takes a value $b(X) \in \mathcal{P}_{\deg(f)}$
as follows.
%
%
\begin{align}
& \mathbb{P} \Big[\mathbf{y}_j(X) \bmod f(X) = b(X) \Big] \nonumber \\
& = \mathbb{P} \Big[[\mathbf{w}_j(X) + \mathbf{e}_j(X)] \bmod f(X) = b(X) \Big] \nonumber  \\
&= \sum_{\mathbf{d}(X) \in \mathcal{P}_{n}} \mathbb{P} \Big[[\mathbf{w}_j(X) + \mathbf{d}(X)] \bmod f(X) = b(X) \Big] \nonumber  \\
&\mbox{~~~~~~~~~~~~~~~~~~~~~~~~~~~~~~~~~~~~~~~~}\mathbb{P}\Big[\mathbf{e}_j(X) = \mathbf{d}(X)\Big] \nonumber  \\
&= \sum_{\mathbf{d}(X) \in \mathcal{P}_{n}} \mathbb{P} \Big[\mathbf{w}_j(X) = \mathbf{d}(X) \bmod f(X) + b(X) \Big] \nonumber  \\
&\mbox{~~~~~~~~~~~~~~~~~~~~~~~~~~~~~~~~~~~~~~~~}\mathbb{P}\Big[\mathbf{e}_j(X) = \mathbf{d}(X)\Big] \nonumber \\
&\stackrel{(a)}{=} \sum_{\mathbf{d}(X) \in \mathcal{P}_{n}} \frac{1}{2^{\deg(f)}} \mathbb{P} \Big[\mathbf{e}_j(X) = \mathbf{d}(X)\Big] \nonumber \\
&= \frac{1}{2^{\deg(f)}} \sum_{\mathbf{d}(X) \in \mathcal{P}_{n}} \mathbb{P} \Big[\mathbf{e}_j(X) = \mathbf{d}(X)\Big] = \frac{1}{2^{\deg(f)}}.
\label{Eqn_yj_uniform}
\end{align}
%
%
where the equality in $(a)$ is obtained since $[\mathbf{d}(X) \bmod f(X) + b(X)]$ is polynomial in $\mathcal{P}_{\deg(f)}$ and 
$\mathbf{w}_j(X) \bmod f(X)$ takes any value in $\mathcal{P}_{\deg(f)}$ with probability $1/2^{\deg(f)}$.
From (\ref{Eqn_yj_uniform}), $\mathbf{y}_j(X) \bmod f(X)$ follows the uniform distribution and the proof is complete.
\end{IEEEproof}

We now consider the case when $\mathbf{w}_j(X) \bmod f(X)$ follows the restricted uniform distribution. 
Let us first consider an example distribution of $\mathbf{y}_j(X) \bmod f(X)$
when $\mathbf{w}_j(X) \bmod f(X)$ follows the restricted uniform distribution.
\begin{example}
\label{Example_type3_noisy}
Suppose code $C(n_0,g_0)$ with $n_0=15$ and $g_0(X) = (X^4+X+1)(X^4+X^3+1)$ is used at the transmitter.
For $n = 10$, suppose $n$-bit vector $\mathbf{w}_j$ is formed by the initial $n = 10$ bits of a codeword in $C(n_0,g_0)$.
For a factor $f(X) = X^4+X^3+X^2+X+1$ of $X^{10}+1$, the distributions of $\mathbf{w}_j(X) \bmod f(X)$ and $\mathbf{y}_j(X) \bmod f(X)$
are shown in \Fig~\ref{Figure_example_type3_n10}(a) and (b) respectively.
It can be seen that,  $\mathbf{w}_j(X) \bmod f(X)$ follows the restricted uniform distribution but the distribution
of $\mathbf{y}_j(X) \bmod f(X)$ is neither uniform nor restricted uniform.
\hfill $\square$
\end{example}
%

%
\begin{figure}[t]

\begin{center}
 
    \begin{tikzpicture}[scale=0.7]

	  \draw [-] [thick] (-0.5,0) -- (11,0);
	  
	  \draw [-] [densely dotted] (0,1.3) -- (9,1.3);	  
	  
	  \draw [black, fill=black](0,1.3) circle (0.05cm);
	  \draw [black, fill=black](2.8,1.3) circle (0.05cm);
	  \draw [black, fill=black](6.3,1.3) circle (0.05cm);	  
	  \draw [black, fill=black](9.1,1.3) circle (0.05cm); 
	  
	  \draw [black, fill=black](0,0) circle (0.025cm);
	  \draw [black, fill=black](0.7,0) circle (0.025cm);
	  \draw [black, fill=black](1.4,0) circle (0.025cm);
	  \draw [black, fill=black](2.1,0) circle (0.025cm);
	  \draw [black, fill=black](2.8,0) circle (0.025cm);
	  \draw [black, fill=black](3.5,0) circle (0.025cm);
	  \draw [black, fill=black](4.2,0) circle (0.025cm);
	  \draw [black, fill=black](4.9,0) circle (0.025cm);
	  \draw [black, fill=black](5.6,0) circle (0.025cm);
	  \draw [black, fill=black](6.3,0) circle (0.025cm);	  
	  \draw [black, fill=black](7,0) circle (0.025cm);
	  \draw [black, fill=black](7.7,0) circle (0.025cm);
	  \draw [black, fill=black](8.4,0) circle (0.025cm);
	  \draw [black, fill=black](9.1,0) circle (0.025cm);
	  \draw [black, fill=black](9.8,0) circle (0.025cm);
	  \draw [black, fill=black](10.5,0) circle (0.025cm);	  	  

	  \draw [-,ultra thick] (0,0) -- (0,1.3);
	  \draw [-,ultra thick] (2.8,0) -- (2.8,1.3);
	  \draw [-,ultra thick] (6.3,0) -- (6.3,1.3);
	  \draw [-,ultra thick] (9.1,0) -- (9.1,1.3);	  

	  \node [below] at (0,0) {\footnotesize $0$};
	  \node [below] at (0.7,0) {\footnotesize $1$};
	  \node [below] at (1.4,0) {\footnotesize $X$};
	  \node [below] at (2.8,0) {\footnotesize $X^2$};
	  \node [below] at (6.3,0) {\footnotesize $X^3+1$};	  
	  \node [below] at (9.1,0) {\footnotesize $X^3+X^2+1$};	  	  

	  \node [left] at (0,1.3) {\footnotesize $0.25$};
	  \draw [-] (0-0.1,1.3) -- (0+0.1,1.3);	  	  
	  \node [below] at (5.3,-1) {(a) Distribution of $\mathbf{w}_j(X) \bmod f(X)$};

	  \draw [-] [thick] (-0.5,0-3.6) -- (11,0-3.6);
	  
	  \draw [-] [densely dotted] (0,1-3.6) -- (10,1-3.6);
	  
	  \draw [black, fill=black](0,1-3.6) circle (0.05cm);
	  \draw [black, fill=black](0.7,0.42-3.6) circle (0.05cm);
	  \draw [black, fill=black](1.4,0.42-3.6) circle (0.05cm);
	  \draw [black, fill=black](2.1,0.18-3.6) circle (0.05cm);
	  \draw [black, fill=black](2.8,1-3.6) circle (0.05cm);
	  \draw [black, fill=black](3.5,0.42-3.6) circle (0.05cm);
	  \draw [black, fill=black](4.2,0.42-3.6) circle (0.05cm);
	  \draw [black, fill=black](4.9,0.18-3.6) circle (0.05cm);
	  \draw [black, fill=black](5.6,0.42-3.6) circle (0.05cm);
	  \draw [black, fill=black](6.3,1-3.6) circle (0.05cm);	  
	  \draw [black, fill=black](7,0.18-3.6) circle (0.05cm);
	  \draw [black, fill=black](7.7,0.42-3.6) circle (0.05cm);
	  \draw [black, fill=black](8.4,0.42-3.6) circle (0.05cm);
	  \draw [black, fill=black](9.1,1-3.6) circle (0.05cm);
	  \draw [black, fill=black](9.8,0.18-3.6) circle (0.05cm);
	  \draw [black, fill=black](10.5,0.42-3.6) circle (0.05cm);

	  \draw [black, fill=black](0,0-3.6) circle (0.025cm);
	  \draw [black, fill=black](0.7,0-3.6) circle (0.025cm);
	  \draw [black, fill=black](1.4,0-3.6) circle (0.025cm);
	  \draw [black, fill=black](2.1,0-3.6) circle (0.025cm);
	  \draw [black, fill=black](2.8,0-3.6) circle (0.025cm);
	  \draw [black, fill=black](3.5,0-3.6) circle (0.025cm);
	  \draw [black, fill=black](4.2,0-3.6) circle (0.025cm);
	  \draw [black, fill=black](4.9,0-3.6) circle (0.025cm);
	  \draw [black, fill=black](5.6,0-3.6) circle (0.025cm);
	  \draw [black, fill=black](6.3,0-3.6) circle (0.025cm);	  
	  \draw [black, fill=black](7,0-3.6) circle (0.025cm);
	  \draw [black, fill=black](7.7,0-3.6) circle (0.025cm);
	  \draw [black, fill=black](8.4,0-3.6) circle (0.025cm);
	  \draw [black, fill=black](9.1,0-3.6) circle (0.025cm);
	  \draw [black, fill=black](9.8,0-3.6) circle (0.025cm);
	  \draw [black, fill=black](10.5,0-3.6) circle (0.025cm);	  	  

	  \draw [-,ultra thick] (0,0-3.6) -- (0,1-3.6);
	  
	  \draw [-,ultra thick] (0.7,0-3.6) -- (0.7,0.42-3.6);	  
	  \draw [-,ultra thick] (1.4,0-3.6) -- (1.4,0.42-3.6);	  	  
	  \draw [-,ultra thick] (2.1,0-3.6) -- (2.1,0.18-3.6);
	  
	  \draw [-,ultra thick] (2.8,0-3.6) -- (2.8,1-3.6);
	  
	  \draw [-,ultra thick] (3.5,0-3.6) -- (3.5,0.42-3.6);	  
	  \draw [-,ultra thick] (4.2,0-3.6) -- (4.2,0.42-3.6);	  
	  \draw [-,ultra thick] (4.9,0-3.6) -- (4.9,0.18-3.6);	  
	  \draw [-,ultra thick] (5.6,0-3.6) -- (5.6,0.42-3.6);	  
	  
	  \draw [-,ultra thick] (6.3,0-3.6) -- (6.3,1-3.6);
	  
	  \draw [-,ultra thick] (7,0-3.6) -- (7,0.18-3.6);	  	  
	  \draw [-,ultra thick] (7.7,0-3.6) -- (7.7,0.42-3.6);	  
	  \draw [-,ultra thick] (8.4,0-3.6) -- (8.4,0.42-3.6);	  	  
	  
	  \draw [-,ultra thick] (9.1,0-3.6) -- (9.1,1-3.6);	  
	  
	  \draw [-,ultra thick] (9.8,0-3.6) -- (9.8,0.18-3.6);	  	  	  
	  \draw [-,ultra thick] (10.5,0-3.6) -- (10.5,0.42-3.6);

	  \node [below] at (0,0-3.6) {\footnotesize $0$};
	  \node [below] at (0.7,0-3.6) {\footnotesize $1$};
	  \node [below] at (1.4,0-3.6) {\footnotesize $X$};
	  \node [below] at (2.8,0-3.6) {\footnotesize $X^2$};
	  \node [below] at (6.3,0-3.6) {\footnotesize $X^3+1$};	  
	  \node [below] at (9.1,0-3.6) {\footnotesize $X^3+X^2+1$};	  	  

	  \node [left] at (0,1-3.6) {\footnotesize $0.125$};
	  \draw [-] (0-0.1,1-3.6) -- (0+0.1,1-3.6);	  	  
	  \node [below] at (5.3,-1-3.6) {(b) Distribution of $\mathbf{y}_j(X) \bmod f(X)$};

    \end{tikzpicture}
\end{center}

\caption{The distributions of $\mathbf{w}_j(X) \bmod f(X)$ and $\mathbf{y}_j(X) \bmod f(X)$ are illustrated when
$\mathbf{w}_j(X)$ is formed by the initial $10$-bits of a codeword in code 
$C(15,g_0)$ with $g_0(X) = (X^4+X+1)(X^4+X^3+1)$ and $f(X) = X^4+X^3+X^2+X+1$.}
\label{Figure_example_type3_n10}
\end{figure}
%

Example~\ref{Example_type3_noisy} suggests that, when $\mathbf{w}_j(X) \bmod f(X)$ follows the restricted uniform distribution, 
the distribution of $\mathbf{y}_j(X) \bmod f(X)$ need not be uniform or restricted uniform.
Let $\mathcal{S}$ be the support set of $\mathbf{w}_j(X) \bmod f(X)$.
In Example~\ref{Example_type3_noisy}, the support set of $\mathbf{w}_j(X) \bmod f(X)$ is $\mathcal{S} = \{ 0, X^2, X^3+1, X^3+X^2+1\}$ 
(see \Fig~\ref{Figure_example_type3_n10}(a)).
From the definition of the restricted uniform distribution, for any $a(X) \in \mathcal{S}$, 
\begin{align}
\mathbb{P}\Big[\mathbf{w}_j(X) \bmod f(X) = a(X)\Big] = \frac{1}{|\mathcal{S}|}.
\label{Eqn_wj_temp}
\end{align}
The probability that $\mathbf{y}_j(X) \bmod f(X)$ takes the value $b(X)\in \mathcal{P}_{\deg(f)}$ is given by,
\begin{align}
\mathbb{P}\Big[& \mathbf{y}_j(X) \bmod f(X) = b(X)\Big] \nonumber \\
&= \mathbb{P}\Big[ [\mathbf{w}_j(X) + \mathbf{e}_j(X)] \bmod f(X) = b(X) \Big] \nonumber \\
&= \mathbb{P}\Big[ \mathbf{e}_j(X) \bmod f(X) = [\mathbf{w}_j(X) \bmod f(X)] + b(X) \Big] \nonumber \\
&\stackrel{(a)}{=} \sum_{a(X) \in \mathcal{S}} \mathbb{P}\Big[ \mathbf{e}_j(X) \bmod f(X) = a(X) + b(X) \Big] \nonumber \\
&\mbox{~~~~~~~~~~~~~~~~~~~~~~~~~} \mathbb{P}\Big[\mathbf{w}_j(X) \bmod f(X) = a(X)\Big] \nonumber \\
&= \frac{1}{|\mathcal{S}|} \sum_{a(X) \in \mathcal{S}} \mathbb{P}\Big[ \mathbf{e}_j(X) \bmod f(X) = a(X) + b(X) \Big],
\label{Eqn_wj_type3_noisy}
\end{align}
where the equality in $(a)$ is obtained by conditioning over the support set $\mathcal{S}$ of $\mathbf{w}_j(X) \bmod f(X)$ and
the last equality is obtained from (\ref{Eqn_wj_temp}).
For $a_1(X), a_2(X) \in \mathcal{S}$ since $a_1(X) + a_2(X) \in \mathcal{S}$,
from (\ref{Eqn_wj_type3_noisy}) we get
\begin{align}
& \mathbb{P}\Big[\mathbf{y}_j(X) \bmod f(X) = a_1(X)\Big] \nonumber \\
&\mbox{~~~~~~~~~~~~~~~}= \mathbb{P}\Big[\mathbf{y}_j(X) \bmod f(X) = a_2(X)\Big].
\label{Eqn_yj_in_S_equal}
\end{align}
From (\ref{Eqn_yj_in_S_equal}), $\mathbf{y}_j(X) \bmod f(X)$ takes any value in $\mathcal{S}$ with the equal probability.
In Example~\ref{Example_type3_noisy}, it can be seen that the probability of observing any two syndromes in $\mathcal{S}$ is the same.
In \Fig~\ref{Figure_example_type3_n10}(b), $\mathbf{y}_j(X) \bmod f(X)$ takes the two values $X^2$ and $X^3+X^2+1$ in $\mathcal{S}$
with equal probability.
However, for calculating the value of $\mathbb{P}[\mathbf{y}_j(X) \bmod f(X) = b(X)]$ for any $b(X) \in \mathcal{P}_{\deg(f)}$
would require the knowledge of the support set $\mathcal{S}$ and the coset weight distribution of code $C(n,f)$ (see (\ref{Eqn_wj_type3_noisy})). 
Since finding the coset weight distribution is NP-hard and the knowledge of the support set $\mathcal{S}$ 
would require the knowledge of the unknown true code $C(n_0,g_0)$,
finding the value of $\mathbb{P}[\mathbf{y}_j(X) \bmod f(X) = b(X)]$ is in general computationally intractable.
Thus finding the distribution of $\mathbf{y}_j(X) \bmod f(X)$ when $\mathbf{w}_j(X) \bmod f(X)$ follows
the restricted uniform distribution is computationally intractable.

%
\section{Application to blind reconstruction of cyclic codes}
\label{Section_Blind_Reconstruction_cyclic_structure}

In the literature, Yardi et al.~\cite{TCOMM_2016} and Zhou et~al.~\cite{Zhou2013_Entropy_new, Zhou2013_Entropy} have proposed
blind reconstruction methods when both the length of the cyclic code and the synchronization of the received data are not known. 
In this section, we provide a theoretical analysis of these methods.

\subsection{A theoretical analysis of the blind reconstruction method proposed in \cite{TCOMM_2016}}

Yardi et al.~have proposed the zero syndrome distribution based method for blind reconstruction~\cite{TCOMM_2016}.
In this method, authors make use of the zero syndromes of the received polynomials.
Suppose $r_j(X) = \mathbf{y}_j(X) \bmod f(X)$ for $j = 1,2,\ldots,M$.
They proved that, for a given $n$, $s$, and $f(X)$ there are either of the following two cases (see Theorem~1 of~\cite{TCOMM_2016}). 
\begin{enumerate} [(1)]
\item When $n = ln_0$ such that $l \in \mathbb{N}$, $s = s_0$, and $f(X)$ is a factor of $g_0(X)$,
\begin{align*}
\mathbb{P}[r_j(X) = 0] = P(C(n,f)),
\end{align*}
for $j = 1, 2, \ldots, M$ and $P(C(n,f))$ is defined as,
\begin{align}
P(C(n,f)) \coloneqq \sum_{i=0}^n A_i p^i (1-p)^{n-i}
\label{Eqn_PCf_definition}
\end{align}
where $\{ A_0, A_1, \cdots, A_n\}$ is the weight distribution of $C(n,f)$.
\item When either $n \neq ln_0$ or $s \neq s_0$ or $f(X)$ is a not factor of $g_0(X)$,
\begin{align*}
\mathbb{P}[r_j(X) = 0] < P(C(n,f)), 
%
\end{align*}
for $j = 1, 2, \ldots, M$.
\end{enumerate}

Using (1) and (2), they formulated and solved the blind reconstruction problem via hypothesis testing problem given by,
\begin{equation}
\begin{aligned}
H_0 &: \mathbb{I}_{\{{r}_j(X)=0\}} \sim \mbox{Bernoulli} \Big( P(C(n,f)) \Big) \\
H_1 &: \mathbb{I}_{\{{r}_j(X)=0\}} \sim \mbox{Bernoulli} \Big( P_j \Big) \mbox{ s.t. } P_j < P(C(n,f)), 
\end{aligned}
\label{Equation_Hypothesis_testing_structure}
\end{equation}
where $j = 1, 2, \ldots, M$ and $\mathbb{I}_{\{{r}_j(X)=0\}}$ is the indicator random variable for 
the event $r_j(X) = 0$.
For analyzing the performance of this method, 
one needs to analyze the performance of the hypothesis testing in (\ref{Equation_Hypothesis_testing_structure}).
The performance of the hypothesis testing can be characterized 
using the Kullback-Leibler (KL) divergence between the two distributions~\cite[Ch.~11]{ThomasCover2006}.
However in (\ref{Equation_Hypothesis_testing_structure}), the distribution under hypothesis $H_1$ is not known in general.
From Theorem~\ref{Theorem_wj_yj_equally_likely}, when $\mathbf{w}_j(X) \bmod f(X)$ follows the uniform
distribution, $\mathbf{y}_j(X) \bmod f(X)$ also follows the uniform distribution and hence in this case under hypothesis $H_1$ we have,
\begin{align}
H_1:\mathbb{I}_{\{{r}_j(X)=0\}} \sim \mbox{Bernoulli} \bigg(\frac{1}{2^{\deg(f)}} \bigg).
\label{Equation_equally}
\end{align}

When $\mathbf{w}_j(X) \bmod f(X)$ follows the restricted uniform distribution,
due to the reasons mentioned in Section~\ref{Section_cyclic_structure_noise_affected}, 
characterizing the distribution of $r_j(X) = \mathbf{y}_j(X) \bmod f(X)$
is computationally intractable.
Hence in the following theorem we provide an upper bound on $\mathbb{P}[r_j(X) = 0]$ which is strictly
less than $P(C(n,f))$.

\begin{theorem}
\label{Theorem_lower_bound_H1_PCnf}
Let $\mathbf{y}_1(X), \mathbf{y}_2(X), \ldots, \mathbf{y}_M(X)$ be the sequence of received polynomials
for an assumed length $n$ and synchronization (see Section~\ref{Section_System_model}).
For a factor $f(X)$ of $X^n+1$, suppose $r_j(X) = \mathbf{y}_j(X) \bmod f(X)$, for $j = 1,2,\ldots,M$.
When either the assumed length $n$ is not a multiple of the correct length $n_0$ or synchronization is not correct or
$f(X)$ is not a factor of the generator polynomial $g_0(X)$ of the code $C(n_0,g_0)$ used at the transmitter,
\begin{align}
\mathbb{P} \big[r_j(X) = 0\big] \leq \mathcal{P}\big(C(n,f)\big) \left( \frac{\lambda + 1}{2} \right),
\label{Eqn_PCnf_UB_thm_eqn}
\end{align}
where the expression for $\mathcal{P}(C(n,f))$ is given in (\ref{Eqn_PCf_definition}) 
and $\lambda$ is defined as follows
\begin{align}
\lambda \coloneqq \frac{1-(1-2p)^{n-\deg(f)+1}}{1+(1-2p)^{n-\deg(f)+1}}.
\end{align}
\end{theorem}
\begin{IEEEproof}
The proof is given in Appendix~F. 
\end{IEEEproof}
%
Using Theorem~\ref{Theorem_lower_bound_H1_PCnf}, we now find a lower bound on the KL-divergence between
the two distributions in (\ref{Equation_Hypothesis_testing_structure}) as follows.
Let $P$ and $Q$ denote the pmf of $\mathbb{I}_{\{{r}_j(X)=0\}}$ under hypothesis $H_0$ and $H_1$
respectively. 
Suppose $P \coloneqq \begin{bmatrix} p_0 & p_1 \end{bmatrix}$ and $Q \coloneqq \begin{bmatrix} q_0 & q_1 \end{bmatrix}$, 
where $p_0$ and $q_0$ are the probabilities of observing the all-zero syndrome under hypotheses $H_0$ and $H_1$ respectively.
A lower bound on the KL-divergence $D_{KL}(P,Q)$ between distributions $P$ and $Q$ is given by~\cite[Sec.~11.6]{ThomasCover2006},
\begin{align}
D_{KL}(P,Q) \geq \frac{1}{2 \ln 2} \Big( |p_0-q_0| + |p_1-q_1| \Big)^2.
\label{Eqn_LB_D_PQ}
\end{align}
Substituting $p_1=1-p_0$ and $q_1=1-q_0$ in (\ref{Eqn_LB_D_PQ}) we get,
\begin{align}
D_{KL}(P,Q) &\geq \frac{1}{2 \ln 2} \Big( |p_0-q_0| + |(1-p_0)-(1-q_0)| \Big)^2  \\
	    &= \frac{1}{2 \ln 2} \Big( |p_0-q_0| + |q_0-p_0| \Big)^2  \\
	    &\stackrel{(a)}{=} \frac{1}{2 \ln 2} \Big( 2|p_0-q_0| \Big)^2 \\
	    &= \frac{2}{\ln 2} \big(p_0-q_0\big)^2, 	    
\label{Eqn_LB_D_PQ2}
\end{align}
where the equality in $(a)$ is obtained since $|p_0-q_0| = |q_0-p_0|$ and the last equality
is obtained since $|p_0-q_0|^2 = (p_0-q_0)^2$.
From Theorem~\ref{Theorem_lower_bound_H1_PCnf} we have, $q_0 \leq p_0 (\lambda + 1)/2$ 
and substituting this in (\ref{Eqn_LB_D_PQ2}) we get,
\begin{align}
D_{KL}(P,Q) &\geq \frac{2}{\ln 2} \bigg[p_0 - \left(\frac{p_0 (\lambda+1) }{2} \right)  \bigg]^2,\\
	    &= \frac{2}{\ln 2} \bigg[p_0 \left(\frac{ 1-\lambda }{2} \right)  \bigg]^2\\
	    &= \frac{2}{\ln 2} \left(\frac{ 1-\lambda }{2} \right)^2 \Big( P(C(n,f)) \Big)^2,	    
\label{Eqn_LB_D_PQ_final}
\end{align}
where the last equality is obtained by substituting $p_0 = P(C(n,f))$.
From (\ref{Eqn_LB_D_PQ_final}), we obtain
a lower bound on the KL-divergence between two distributions in (\ref{Equation_Hypothesis_testing_structure}).

To summarize, depending on whether $\mathbf{w}_j(X) \bmod f(X)$ follows the uniform or the restricted uniform distribution, 
we can characterize the distribution of $\mathbb{I}_{\{{r}_j(X)=0\}}$ using (\ref{Equation_equally}) and (\ref{Eqn_PCnf_UB_thm_eqn}),
which is required for analyzing the performance of the hypothesis testing of (\ref{Equation_Hypothesis_testing_structure}) in 
the zero syndrome distribution based method.

\subsection{A theoretical analysis of the blind reconstruction method proposed in \cite{Zhou2013_Entropy_new}}

Zhou et al.~have proposed the factor-entropy based method for blind reconstruction of binary cyclic codes~\cite{Zhou2013_Entropy_new}.
The basic idea of this method is as follows.
Suppose $H$ is a parity check matrix of the code $C(n,f)$ generated by a factor $f(X)$ of $X^n+1$.
They consider the sequence received vectors $\mathbf{y}_1, \mathbf{y}_2, \ldots, \mathbf{y}_M$
and find the inner product of each $\mathbf{y}_j$ with $H$ given by,
\begin{align}
\mathbf{y}_j H^T = \Big[ r_{j,0} \mbox{~~} r_{j,1} \mbox{~~} \ldots \mbox{~~} r_{j,\deg(f)-1} \Big].
\end{align}
They define the \textit{mean value of probability of zero syndrome} $P(\mathbf{y}_j,f)$ as,
\begin{align}
P(\mathbf{y}_j,f) \coloneqq \frac{1}{\deg(f)} \sum_{l=0}^{\deg(f)-1} \mathbb{P}[r_{j,l} = 0].
\label{Eqn_chinese_Pjf_expression}
\end{align}
For blind reconstruction they assume that, when either $n$ or $s$ is incorrect, 
\begin{align}
P(\mathbf{y}_j,f_1) = P(\mathbf{y}_j,f_2),
\label{Eqn_chinese_equally_f1_f2_Pyjf}
\end{align}
where $f_1(X)$ and $f_2(X)$ are any two factors of $X^n+1$. They further assume that when $n = n_0$ and $s=s_0$
the assumption in (\ref{Eqn_chinese_equally_f1_f2_Pyjf}) is not valid.
The correct length and synchronization are distinguished from any incorrect ones using this assumption.

In this section, we verify the validity of the assumption in (\ref{Eqn_chinese_equally_f1_f2_Pyjf}).
We next illustrate an example situation where for an incorrect $s$, $P(\mathbf{y}_j,f_1) \neq P(\mathbf{y}_j,f_2)$
which implies that the assumption in (\ref{Eqn_chinese_equally_f1_f2_Pyjf}) is not correct.
\begin{example}
\label{Example_chinese_counterEx}
Suppose the cyclic code $C(n_0,g_0)$ with $n_0=7$ and $g(X)=X^3+X+1$ is used for the communication.
Let us assume that $s_0=0$ is the correct synchronization.
For $n=7$ and $s=1$, the values of $P(\mathbf{y}_j,f)$
for all possible factors of $X^{7}+1$ are provided in Table~\ref{Table_chinese_counterex}, for $j = 1,2,\ldots,M$.
\begin{table}[htbp]
\begin{center}
    \begin{tabular}{| c | c | c | c | }
    \hline
    $f(X)$   	&  $P(\mathbf{y}_j,f)$ 	&  $P(\mathbf{y}_j,f)$ 	&  $P(\mathbf{y}_j,f)$ 	\\ 
         	&  for $p=0$ 		&  for $p=0.01$  	&  for $p=0.05$ 	\\ \hline    
    $X+1$ 	&  $0.5$ 		&  $0.5$ 		&  $0.5$ 		\\ \hline
    $X^3+X+1$ 	&  $0.8334$ 		&  $0.8076$		&  $0.7184$		\\ \hline
    $X^3+X^2+1$ &  $0.5$ 		&  $0.5$ 		&  $0.5$ 		\\ \hline
    \end{tabular}
\end{center}
\caption{The values of $P(\mathbf{y}_j,f)$ for all possible factors of $X^{7}+1$ are illustrated when $n=7$ and $s=1$.}
\label{Table_chinese_counterex}
\end{table}
%
%
It can be seen that, for the chosen incorrect parameters, the assumption in (\ref{Eqn_chinese_equally_f1_f2_Pyjf})
is not valid.
In this example, note that the assumed length $n$ was correct, $f(X)$ was a factor of $g_0(X)$, but the assumed
synchronization was not correct. In general, when $n=n_0$ but $s\neq s_0$, the  assumption
in (\ref{Eqn_chinese_equally_f1_f2_Pyjf}) need not be true.
\hfill $\square$
\end{example}

Though the assumption in (\ref{Eqn_chinese_equally_f1_f2_Pyjf}) is not valid always, in the following
theorem we prove that, the assumption in (\ref{Eqn_chinese_equally_f1_f2_Pyjf}) is true when 
$C(n_0,g_0)$ not degenerate and the assumed length $n < n_0$.

\begin{theorem}
\label{Theorem_chinese}
Suppose the true code $C(n_0,g_0)$ used at the transmitter is not degenerate.
For an assumed length $n$ and synchronization $s$, let $\mathbf{y}_j$ and $\mathbf{w}_j$ be the $j$th noise-affected and 
noise-free $n$-bits vectors respectively, for $j =1,2,\ldots,M$.
For a factor $f(X)$ of $X^n+1$, let $P(\mathbf{y}_j,f)$ be as defined in (\ref{Eqn_chinese_Pjf_expression}).
If $n < n_0$, we have $P(\mathbf{y}_j,f_1) = P(\mathbf{y}_j,f_2)$, where $f_1(X)$ and $f_2(X)$ are
any two factors of $X^n+1$.
\end{theorem}
\begin{IEEEproof}
The proof is given in Appendix~G. 
\end{IEEEproof}
%

\subsection{A theoretical analysis of blind reconstruction method proposed in \cite{Zhou2013_Entropy}}

In \cite{Zhou2013_Entropy},  Zhou et al.~have proposed the root-entropy based method for blind reconstruction.
In this method, for an assumed length $n$ and synchronization $s$, 
authors consider the received polynomials $\mathbf{y}_1(X), \mathbf{y}_2(X), \ldots, \mathbf{y}_M(X)$.
They find an empirical probability of each root $\beta$ of $X^n+1$ being a root of the received polynomials.
They assume that, when either $n$ or $s$ is incorrect, all possible roots of $X^n+1$ are equally likely to be roots
of the received polynomials, i.e., for any two roots $\beta_1$ and $\beta_2$ of $X^n+1$,
\begin{align}
\mathbb{P} \Big[ \beta_1 \mbox{ is a root of } \mathbf{y}_j(X) \Big] = \mathbb{P} \Big[ \beta_2 \mbox{ is a root of } \mathbf{y}_j(X) \Big],
\label{Eqn_root_entropy_assumption}
\end{align}
for $j = 1,2,\ldots,M$.
In Example~\ref{Example_chinese_counterEx_root}, we provide an example situation when this assumption
is not true.
%
\begin{example}
\label{Example_chinese_counterEx_root}
Suppose code $C(n_0,g_0)$ with $n_0=15$ and $g_0(X) = (X^4+X^3+1)(X^4+X^3+X^2+X+1)(X+1)$ 
is used at the transmitter and $s_0=0$ is the correct synchronization of the received sequence.
For an assumed length $n=7$ and synchronization $s=0$,
the first $n$-bit received vector $\mathbf{y}_1$ will be $\mathbf{y}_1 = \mathbf{w}_1 + \mathbf{e}_1$, where
$\mathbf{w}_1$ is formed by the initial $7$ bits of a codeword in code $C(15,g_0)$.
Consider two roots $\beta_1$ and $\beta_2$ of $X^7+1$, whose minimal polynomials are
$f_1(X) = X+1$ and $f_2(X) = X^3+X+1$ respectively.
It is known that, $\beta_i$ is a root of $\mathbf{y}_1(X)$ if and only if 
the minimal polynomial $f_i(X)$ of $\beta_i$ is a factor of $\mathbf{y}_1(X)$, for $i=1,2$~\cite[Sec.~2.2]{Lidl86}.
Thus the probability of a given $\beta_i$ is a root of $\mathbf{y}_1(X)$ is the same as that of 
the probability that $f_i(X)$ is a factor of $\mathbf{y}_1(X)$.
For a factor $f_i(X)$ of $X^7+1$, the probability that $f_i(X)$ is a factor of $\mathbf{y}_1(X)$ can be found 
by conditioning over all possible codewords in $C(15,g_0)$.
For two factors $X^3+X+1$ and $X+1$ of $X^7+1$, it can be verified that for any value of crossover probability $p$,
\begin{align*}
\mathbb{P} \Big[ \beta_1 \mbox{ is a root of } \mathbf{y}_1(X) \Big] &= \frac{1}{2^{\deg(f)}} = 0.125 \\
\mathbb{P} \Big[ \beta_2 \mbox{ is a root of } \mathbf{y}_1(X) \Big] &= \frac{1}{2^{\deg(f)}} = 0.5.
\end{align*}
It can be seen that, the assumption in (\ref{Eqn_root_entropy_assumption}) is not valid for $\mathbf{y}_1(X)$.
\hfill $\square$
\end{example}
%


\section{Conclusion}
\label{Section_conclusion}

In this paper, we analyzed the syndrome distribution of the noise-free and
noise-affected received sequence.
For the noise-free case, we completely characterized the syndrome distribution of the received sequence.
We proved that the distribution of syndrome of any noise-free received polynomial with respect to a candidate polynomial $f(X)$
is degenerate if and only if the assumed length is an integer multiple of the correct length, 
assumed synchronization is correct, and $f(X)$ is a factor of the generator polynomial of the true code.
We proved that, in all the remaining cases the distribution can either be
uniform or restricted uniform.
We also provided the conditions under which this distribution will be of either of the type.
For the noise-affected situation
we observed that, while the syndrome distribution could be completely characterized 
for some of the assumed parameters, in general finding this distribution becomes computationally intractable.
Finally, we provided a theoretical analysis of the existing methods available in the literature for blind reconstruction.

\section*{Appendix A: Some properties of linear block codes} 

\begin{lemma}
\label{Lemma_h_not_dual_equally}
Consider a non-trivial linear block code $C(n)$ of length $n$ and dimension $k$.
Every codeword in this code is chosen according to the uniform distribution.
Consider a codeword $\mathbf{v} \in C(n)$ 
and a vector $\mathbf{h} \in \mathbb{F}_2^n$. 
Then the inner product $\mathbf{v} \mathbf{h}^{T}$
is equally likely to be zero or one if and only if $\mathbf{h} \notin C^{\perp}(n)$.
\end{lemma}
\begin{IEEEproof}
Suppose the inner product $\mathbf{v} \mathbf{h}^{T}$ is equally likely
to be zero or one. If $\mathbf{h} \in C^{\perp}(n)$, then the inner product
$\mathbf{v} \mathbf{h}^{T}$ will always be zero, which is a contradiction.
This implies that $\mathbf{h} \notin C^{\perp}(n)$ and the proof is complete.

We now prove the converse. Suppose $\mathbf{h} \notin C^{\perp}(n)$.
Suppose $\langle C^{\perp}(n), \mathbf{h} \rangle$ denotes the subspace spanned by a set of linearly independent vectors of 
$C^{\perp}(n)$ and $\mathbf{h}$. 
In this case we have $C^{\perp}(n) \subseteq \langle C^{\perp}(n), \mathbf{h} \rangle$ and this implies that,
\begin{align}
\Big\langle C^{\perp}(n), \mathbf{h} \Big \rangle^{\perp} \subseteq \Big(C^{\perp}(n)\Big)^{\perp} = C(n). 
\label{Eqn_lemma1_1}
\end{align}
Since the dimension of $C(n)$ is $k$,
the dimensions of $C^{\perp}(n)$ and $\langle C^{\perp}(n),\mathbf{h} \rangle$ will be $n-k$ and $n-k+1$
respectively. 
The dimension of $\langle C^{\perp}(n), \mathbf{h}\rangle^{\perp}$ is $n - (n-k+1)= k-1$ 
and hence in (\ref{Eqn_lemma1_1}) we have,
\begin{align}
\Big \langle C^{\perp}(n), \mathbf{h} \Big \rangle^{\perp} \mbox{~} \subset \mbox{~} C(n). 
\label{Eqn_lemma1_2}
\end{align}
Since $\mathbf{h} \notin C^{\perp}(n)$ and the dimension of $\langle C^{\perp}(n), \mathbf{h} \rangle^{\perp}$
is exactly one less than the dimension of $C(n)$, 
from (\ref{Eqn_lemma1_2}), the inner product $\mathbf{v} \mathbf{h}^{T}$ will be
zero for exactly $2^{k-1}$ number of codewords in $C(n)$.
Since every $\mathbf{v} \in C(n)$ is chosen according to the uniform distribution,
$\mathbf{v} \mathbf{h}^{T}$ will be equally likely to be zero or one and the proof is complete.
\end{IEEEproof}

%
\begin{lemma}
\label{Lemma_for_sync_equally_likely}
Consider a linear block code $C(n)$ and a vector $\mathbf{h} \notin C^{\perp}(n)$.
For positive integers $d_1$ and $d_2$, let $C_1(d_1)$ and $C_2(d_2)$ be the linear subspaces formed by the
set of prefixes and suffixes of codewords in $C(n)$ of lengths $d_1$ and $d_2$ respectively, where $1 \leq d_1, d_2 < n$
such that $d_1+d_2=n$.
Then either of the following is true. 
\begin{enumerate}[(1)]
\item $\mathbf{h}(0:d_1-1) \notin C_1^{\perp}(d_1)$
\item $\mathbf{h}(d_1:n-1) \notin C_2^{\perp}(d_2)$
\end{enumerate}

\end{lemma}

\begin{IEEEproof}
When $\mathbf{h}(0:d_1-1) \notin C_1^{\perp}(d_1)$, 
condition (1) of the lemma is satisfied and the lemma is trivially true.
Hence we consider the case when $\mathbf{h}(0:d_1-1) \in C_1^{\perp}(d_1)$.
This implies that, for any $\mathbf{v} \in C(n)$, 
\begin{align}
\mathbf{v}(0:d_1-1)\mathbf{h}(0:d_1-1)^{T} = 0.
\label{Eqn_Lemma2_1}
\end{align}
We now prove by contradiction that $\mathbf{h}(d_1:n-1) \notin C_2^{\perp}(d_2)$.
When $\mathbf{h}(d_1:n-1) \in C_2^{\perp}(d_2)$, for any $\mathbf{v} \in C(n)$,
\begin{align}
\mathbf{v}(d_1:n-1) \mathbf{h}(d_1:n-1)^{T} = 0.
\label{Eqn_Lemma2_2}
\end{align}
From (\ref{Eqn_Lemma2_1}) and (\ref{Eqn_Lemma2_2}), $\mathbf{v} \mathbf{h}^{T} $ will always be zero.
Since it is given that $\mathbf{h} \notin C^{\perp}(n)$,
we get a contradiction from Lemma~\ref{Lemma_h_not_dual_equally}.
Hence $\mathbf{h}(d_1:n-1) \notin C_2^{\perp}(d_2)$. 
Thus the condition (2) of the lemma is satisfied and the proof is complete.
\end{IEEEproof}

\begin{lemma}
\label{Lemma_dual_gen_factor_degenerate_pattern}
Consider a cyclic code $C(n,g)$ of length $n$ and generator polynomial $g(X)$.
Let $g^{\perp}(X)$ be the generator polynomial of the dual code of $C(n,g)$.
Then $C(n,g)$ contains a codeword of a degenerate pattern if and only if there exists a 
factor of $g^{\perp}(X)$ whose order is strictly less than $n$ (see Definitions~\ref{Definition_order_fX}, \ref{Definition_degenerate_pattern}).
\end{lemma}
\begin{IEEEproof}
%
%
Suppose $g^{\perp}(X)$ has a factor $f^{\perp}(X)$ of order $n^{\prime}$ such that $1 \leq n^{\prime} < n$.
Let $f(X)$ be the generator polynomial of the dual code of $C(n,f^{\perp})$.
It is known that $C(n,f) \subseteq C(n,g)$~\cite[Sec.~7.4]{Peterson_1996}.  
Since the order of $f^{\perp}(X)$ is strictly less than $n$, $C(n,f)$ will be 
a degenerate code~\cite[Sec.~8.3]{Macwilliams_Sloane_1977} and the proof is complete since $C(n,f) \subseteq C(n,g)$. 

We now prove the converse. Suppose there exists a codeword $\mathbf{v} \in C(n,g)$ of a degenerate pattern, i.e.,
$\mathbf{v}$ can be written as,
\begin{align}
\mathbf{v} = \Big[\underbrace{\mathbf{w} \mbox{~~} \mathbf{w} \mbox{~~} \cdots \mbox{~~} \mathbf{w}}_{l \text{ times}}\Big],
\label{Eqn_v_app_LRR0}
\end{align}
where $l > 1$ and $\mathbf{w}$ is a vector of length $n^{\prime} = n/l$ such that
$\mathbf{w}$ is not a vector of a degenerate pattern.
Since $\mathbf{v}$ is a codeword in a cyclic code, the vector $\mathbf{v}^{(i)}$ obtained by $i$ right cyclic shifts of 
$\mathbf{v}$ will also be a codeword in $C(n,g)$ given by,
\begin{align}
\mathbf{v}^{(i)} = \Big[\underbrace{\mathbf{w}^{(i)} \mbox{~~} \mathbf{w}^{(i)} \mbox{~~} \cdots \mbox{~~} \mathbf{w}^{(i)}}_{l \text{ times}}\Big],
\label{Eqn_v_app_LRR}
\end{align}
where $\mathbf{w}^{(i)}$ is the vector obtained by $i$ right cyclic shifts of $\mathbf{w}$ and $1 \leq i < n$.
From (\ref{Eqn_v_app_LRR0}), vector $\mathbf{v}^{(n^{\prime})}$ obtained by $n^{\prime}$ right cyclic shifts of $\mathbf{v}$
will be equal to $\begin{bmatrix} \mathbf{w} & \mathbf{w} & \ldots & \mathbf{w} \end{bmatrix} = \mathbf{v}$.
Thus the set of codewords $\{ \mathbf{v}^{(1)}, \mathbf{v}^{(2)}, \ldots, \mathbf{v}^{(n^{\prime})} = \mathbf{v} \}$
will be distinct.
It can be easily shown that the subspace spanned by $\{ \mathbf{v}^{(1)}, \mathbf{v}^{(2)}, \ldots, \mathbf{v}^{(n^{\prime})} \}$
is a cyclic code.
Let $f(X)$ be the generator polynomial of this code, denoted by $C(n,f)$.
Observe that every codeword in code $C(n,f)$ is of a degenerate pattern (see (\ref{Eqn_v_app_LRR})), i.e.,
$C(n,f)$ is a degenerate cyclic code such that $C(n,f) \subseteq C(n,g)$.
%
This implies that $C(n,f^{\perp}) \supseteq C(n,g^{\perp})$~\cite{LinCostello2004} and 
$f^{\perp}(X)$ is a factor of $g^{\perp}(X)$.
%

From (\ref{Eqn_v_app_LRR}), the period of the linear recurring sequence given by 
$[\mathbf{w}^{(i)} \mbox{~} \mathbf{w}^{(i)} \mbox{~} \cdots]$ is $n^{\prime}$, which implies that 
$f(X)$ divides $X^{n^{\prime}}+1$~\cite[Sec.~3.1]{Lidl86}. Since $n^{\prime} < n$, the order of $f^{\perp}(X)$
is strictly less than $n$ and the proof is complete.
\end{IEEEproof}
%

\section*{Appendix B: Proof of Proposition~\ref{Proposition_cyclic_structure_no_type1}}
%
In this appendix, we will prove Proposition~\ref{Proposition_cyclic_structure_no_type1}.
We first summarize some properties of syndrome $r(X)$ that will be required to prove this proposition.

\begin{property}
\label{Property_division1}
Let $\mathcal{W}$ be a linear subspace of $\mathbb{F}_2^n$.
Let $\mathbf{w}(X)$ be the polynomial corresponding to $\mathbf{w} \in \mathcal{W}$.
For a polynomial $f(X) \in \mathbb{F}_2[X]$, suppose the syndrome $r(X)$ of $\mathbf{w}(X)$ with respect to $f(X)$ is given by,
\begin{equation}
\begin{aligned}
r(X) &= \mathbf{w}(X) \bmod f(X) \\
&= r_0 + r_1X + \ldots + r_{\deg(f)-1}X^{\deg(f)-1},
%
\end{aligned}
\end{equation}
where each $r_l \in \mathbb{F}_2$, for $l = 0,1,\ldots,\deg(f)-1$.
Then for every coefficient $r_l$ of $r(X)$, there exists a vector $\mathbf{h}_l \in \mathbb{F}_2^n$ such that
\begin{align*}
r_l = \mathbf{w} \mathbf{h}_l^T,
\label{Eqn_r_l_w_h_perp}
\end{align*}
where $l = 0,1,\ldots,\deg(f)-1$.
\hfill $\square$
\end{property}
%
%
\begin{proof}
For polynomial $f(X)$, define the map $L$ acting on $\mathbf{w} \in \mathcal{W}$ as follows,
\begin{align}
L(\mathbf{w}) \coloneqq \mathbf{w}(X) \bmod f(X) = r(X).
\end{align}
It can be seen that $L$ is a linear map.
Let $\mathbf{r}$ be the vector corresponding to $r(X)$, where $\mathbf{r} \in \mathbb{F}_2^{\deg(f)}$. 
Since $\mathbf{w}$ and $\mathbf{r}$ are in one-to-one correspondence with 
$\mathbf{w}(X)$ and $r(X)$ respectively, the linear map $L$ can be given by 
\begin{align}
L : \mathbb{F}_2^n \rightarrow \mathbb{F}_2^{\deg(f)}.
\end{align}
It is known that, corresponding to every linear transformation $L : \mathbb{F}_2^n \rightarrow \mathbb{F}_2^{\deg(f)}$, 
there exists some matrix $A \in \mathbb{F}_2^{n \times \deg(f)}$ associated to it such that
\begin{align}
L (\mathbf{w}) = \mathbf{w} A = \mathbf{r}, 
\label{Eqn_r_l_proof1}
\end{align}
where $\mathbf{w} \in \mathbb{F}_2^n$ and $\mathbf{r} \in \mathbb{F}_2^{\deg(f)}$ are considered as row vectors~\cite[Ch.~4]{Artin_Algebra}.
Suppose matrix $A$ is given by,
\begin{align}
A = \begin{bmatrix} \mathbf{h}_0 & \mathbf{h}_1 & \cdots & \mathbf{h}_{\deg(f)-1} \end{bmatrix},
\label{Eqn_r_l_proof2}
\end{align}
where $\mathbf{h}_l \in \mathbb{F}_2^n$ for $l = 0,1,\ldots,\deg(f)-1$ are the columns of matrix $A$.
From (\ref{Eqn_r_l_proof1}) and (\ref{Eqn_r_l_proof2}) it can be seen that the $l$th coefficient $r_l$ of $\mathbf{r}$
can be written as $\mathbf{w}\mathbf{h}_l^T$ and the proof is complete.
\end{proof}
Let us consider an example to explain this property. 
\begin{example}
\label{Example_propertyE1}
For $n=7$ and $f(X) = X^3+X^2+1$, $r(X)$ is given by
%
%
\begin{align}
r(X) &= \mathbf{w}(X) \bmod f(X) = r_0 + r_1X +r_2X^2 \nonumber \\
     & =(w_0 + w_1 X + \ldots + w_6 X^6) \bmod (X^3+X^2+1) \nonumber \\
     &= \Big[w_0 + w_3 + w_4 + w_5\Big] + \Big[w_1 + w_4 + w_5 + w_6\Big]X \nonumber \\
     &\mbox{~~~~~~~~~~~~~~~} + \Big[w_2 + w_3 + w_4 + w_6\Big]X^2 \nonumber \\
     & = \mathbf{w}\mathbf{h}_0^T + \mathbf{w}\mathbf{h}_1^T X + \mathbf{w}\mathbf{h}_2^T X^2,
%
\label{Eqn_rx_detail}
\end{align}
%
where $\mathbf{h}_0 = [1 \mbox{~} 0 \mbox{~} 0 \mbox{~}1 \mbox{~}1 \mbox{~}1 \mbox{~}0]$, 
$\mathbf{h}_1 = [0 \mbox{~} 1 \mbox{~} 0 \mbox{~} 0 \mbox{~} 1 \mbox{~} 1 \mbox{~} 1 ]$ and
$\mathbf{h}_2 = [0 \mbox{~} 0 \mbox{~} 1 \mbox{~} 1 \mbox{~} 1 \mbox{~} 0 \mbox{~}1 ]$.
%
%
%
\hfill $\square$
\end{example}

%

\begin{property}
\label{Property_division2}
When $f(X)$ is a factor of $X^n+1$, the set of vectors $\{ \mathbf{h}_0, \mathbf{h}_1, \ldots, \mathbf{h}_{\deg(f)-1} \}$ in 
Property~\ref{Property_division1} form a basis for the dual code $C(n,f^{\perp})$
of cyclic code $C(n,f)$.
\hfill $\square$
\end{property}
We first provide an example of this property and then provide a proof.
\begin{example}
In Example~\ref{Example_propertyE1}, $f(X)=X^3+X^2+1$ is a factor of $X^7+1$.
The generator polynomial of the dual code of $C(7,f)$ is $f^{\perp}(X) = (X+1)(X^3+X^2+1)$.
In (\ref{Eqn_rx_detail}), the polynomials corresponding to $\mathbf{h}_0,\mathbf{h}_1$, and $\mathbf{h}_2$ are given by, 
$\mathbf{h}_0(X) = (X+1)f^{\perp}(X)$, $\mathbf{h}_1(X) = X(X+1)f^{\perp}(X)$, and $\mathbf{h}_2(X) = X^2f^{\perp}(X)$ respectively.
It can be seen that $\mathbf{h}_0,\mathbf{h}_1,\mathbf{h}_2$ are linearly independent and form a basis of $C(7,f^{\perp})$.
\hfill $\square$
\end{example}

\textit{Proof of Property~\ref{Property_division2}}:
%
%
For $\mathbf{w}(X) \in \mathcal{P}_n$, it is known that $\mathbf{w}(X) \in C(n,f)$ if and only if 
$r(X) = \mathbf{w}(X) \bmod f(X) = 0$~\cite{LinCostello2004}.
From (\ref{Eqn_r_l_w_h_perp}), $r(X) = 0$ implies that $r_l = \mathbf{w}\mathbf{h}_l^T = 0$, for $l = 0,1,\ldots,\deg(f)-1$.
The inner product $\mathbf{w}\mathbf{h}_l^T = 0$ for every $\mathbf{w} \in C(n,f)$ implies that
$\mathbf{h}_l \in C(n,f^{\perp})$, where $C(n,f^{\perp})$ is the dual code of $C(n,f)$.
Using this, the code $C(n,f)$ is given by,
\begin{align}
C(n,f) = \Big\{ \mathbf{w} \in \mathbb{F}_2^n \Big| \mathbf{w}\mathbf{h}_l^T=0 \mbox{ for } l = 0,1,\ldots,\deg(f)-1 \Big\}.
\label{Eqn_set_hi_Cnf}
\end{align}
%
%
We now prove by contradiction that the set of vectors $\{ \mathbf{h}_0, \mathbf{h}_1, \ldots, \mathbf{h}_{\deg(f)-1} \}$ in (\ref{Eqn_set_hi_Cnf})
are independent which completes the proof of the property. 
Without loss of generality suppose $\mathbf{h}_0$ can be written a linear combination of  $\{ \mathbf{h}_1, \ldots, \mathbf{h}_{\deg(f)-1} \}$ 
given by,
\begin{align}
\mathbf{h}_0 = a_1 \mathbf{h}_1 + a_2 \mathbf{h}_2 + \ldots + a_{\deg(f)-1} \mathbf{h}_{\deg(f)-1},
\label{Eqn_set_hi_Cnf2}
\end{align}
where $a_i \in \mathbb{F}_2$ for $i = 1,2,\ldots,{\deg(f)-1}$.
From (\ref{Eqn_set_hi_Cnf2}), if $\mathbf{w}\mathbf{h}_i^T = 0$ for $i = 1,2,\ldots,{\deg(f)-1}$ 
we get $\mathbf{w}\mathbf{h}_0^T = 0$. 
Using this in (\ref{Eqn_set_hi_Cnf}) we get,
\begin{align}
C(n,f) = \Big\{ \mathbf{w} \in \mathbb{F}_2^n \Big| \mathbf{w}\mathbf{h}_i^T=0 \mbox{ for } i = 1,\ldots,\deg(f)-1 \Big\}.
\label{Eqn_set_hi_Cnf3}
\end{align}
From (\ref{Eqn_set_hi_Cnf3}), the dimension of $C(n,f)$ should be greater than or equal to $n-\deg(f)+1$.
This is a contradiction since the dimension of $C(n,f)$ is equal to $n-\deg(f)$~\cite{LinCostello2004}.
This completes the proof.
\hfill $\blacksquare$

Using Properties~\ref{Property_division1} and \ref{Property_division2}
we now characterize the distribution of 
$r(X) = \mathbf{w}(X) \bmod f(X)$, when $\mathbf{w}(X)$ lies in any linear subspace $\mathcal{W}(n)$ in the following lemma.

\begin{lemma}
\label{Lemma_three_type_distri}
Consider a linear subspace $\mathcal{W}(n)$ of $\mathbb{F}_2^n$.
Suppose every $\mathbf{w}(X) \in \mathcal{W}(n)$ is chosen i.i.d.~according to the uniform distribution.
For a factor $f(X)$ of $X^n+1$, suppose $r(X) = \mathbf{w}(X) \bmod f(X)$.
Then the distribution of the random variable corresponding to $r(X)$ can either be degenerate or uniform
or restricted uniform (see (\ref{Eqn_type_1_distri}), (\ref{Eqn_type_2_distri}), (\ref{Eqn_type_3_distri}), \Fig~\ref{Figure_example_distributions}).
\end{lemma}
\begin{IEEEproof}
Suppose $r(X) = \mathbf{w}(X) \bmod f(X)$ is given by,
\begin{equation}
\begin{aligned}
r(X) &= r_0 +r_1X + \ldots + r_{{\deg(f)}-1}X^{{\deg(f)}-1} \\ 
	      &= \mathbf{w}\mathbf{h}_0^T +\mathbf{w}\mathbf{h}_1^T X + \ldots + \mathbf{w}\mathbf{h}_{{\deg(f)}-1}^TX^{{\deg(f)}-1},
\label{Eqn_r_X}
\end{aligned}
\end{equation}
where the last equality is obtained from Property~\ref{Property_division1} such that each $\mathbf{h}_l \in \mathbb{F}_2^n$
for $l = 0,1,\ldots,\deg(f)-1$.
%
Let $\{ R_0, R_1, \ldots, R_{\deg(f)-1} \}$ be the set of random variables corresponding to 
$\{ r_0, r_1, \ldots, r_{\deg(f)-1} \}$.
For a given $\mathbf{h}_l$ and $\mathcal{W}(n)$ there are two possibilities, either $\mathbf{h}_l \in \mathcal{W}^{\perp}(n)$
or $\mathbf{h}_l \notin \mathcal{W}^{\perp}(n)$, where $\mathcal{W}^{\perp}(n)$ is the dual code of $\mathcal{W}(n)$.
When $\mathbf{h}_l \in \mathcal{W}^{\perp}(n)$, the corresponding $R_l$ is always zero and 
when $\mathbf{h}_l \notin \mathcal{W}^{\perp}(n)$, 
from Lemma~\ref{Lemma_h_not_dual_equally} of Appendix~A,
the corresponding $R_l$ is equally likely to be zero or one.
We now consider various situations for the set of random variables $\{ R_0, R_1, \ldots, R_{\deg(f)-1} \}$.
\begin{enumerate} [(i)]
\item Each $R_l$ for $l = 0, 1, \ldots, \deg(f)-1$ is zero with probability one. 
This implies that in (\ref{Eqn_r_X}), the random variable corresponding to $r(X)$ is zero with probability one, which is  
the degenerate distribution (see \Fig~\ref{Figure_example_distributions} (a), (\ref{Eqn_type_1_distri})).
\item The set of random variables $\{ R_0, R_1, \ldots, R_{\deg(f)-1} \}$ satisfy a linear relation given by
\begin{align}
a_0R_0 + a_1 R_1 + \ldots + a_{\deg(f)-1} R_{\deg(f)-1} = 0,
\label{Eqn_r_i_lin_combi_pre1}
\end{align}
where each $a_l \in \mathbb{F}_2$, for $l = 0, 1, \ldots, \deg(f)-1$.
We consider the case when at least one of the $R_l$ is equally likely to be one or zero, otherwise 
this case will get reduced to case (i).
From (\ref{Eqn_r_i_lin_combi_pre1}), $R_0$ depends on $R_1, R_2, \ldots R_{\deg(f)-1}$.
Thus the random vector $[R_0 \mbox{~} R_1 \mbox{~} \ldots \mbox{~} R_{\deg(f)-1}]$
cannot take all possible $2^{\deg(f)}$ values.
As each $R_l$ is either zero with probability one or equally likely to be zero or one,
in (\ref{Eqn_r_X}) $r(X)$ will follow the restricted uniform distribution (see \Fig~\ref{Figure_example_distributions} (c), (\ref{Eqn_type_3_distri})).
\item The set of random variables $\{ R_0, R_1, \ldots, R_{\deg(f)-1} \}$ are independent.
In this case, when each $R_l$ for $l = 0, 1, \ldots, \deg(f)-1$ is equally likely to be zero or one,
the random variable corresponding to $r(X)$ will take all possible $2^{\deg(f)}$ values with equal probability, which is
the uniform distribution (see \Fig~\ref{Figure_example_distributions} (b), (\ref{Eqn_type_2_distri})).
\end{enumerate}

We now show that the set of random variables $\{ R_0, R_1, \ldots, R_{\deg(f)-1} \}$
satisfies either of the above situations, which completes the proof.
Let us first consider the case when $\mathcal{W}(n)$ is a nontrivial code. 
The case when $\mathcal{W}(n)$ is a trivial code will be considered later in this proof.
For the two codes $\mathcal{W}^{\perp}(n)$ and $C(n,f^{\perp})$ there are the
following four possibilities.
%
\begin{enumerate}[1)]
\item $\mathcal{W}^{\perp}(n) = C(n,f^{\perp})$ 
\item $C(n,f^{\perp}) \subset \mathcal{W}^{\perp}(n)$, where $\subset$ denotes strict subset
\item $\mathcal{W}^{\perp}(n) \subset C(n,f^{\perp})$
\item $\mathcal{W}^{\perp}(n) \nsubseteq C(n,f^{\perp})$ and $C(n,f^{\perp}) \nsubseteq \mathcal{W}^{\perp}(n)$
\end{enumerate}

In cases 1) and 2), we have $C(n,f^{\perp}) \subseteq \mathcal{W}^{\perp}(n)$.
From Property~\ref{Property_division2}, every $\mathbf{h}_l \in C(n,f^{\perp})$ 
and hence we have $\mathbf{h}_l \in \mathcal{W}^{\perp}(n)$,
for $l = 0, 1, \ldots, \deg(f)-1$ (see (\ref{Eqn_r_X})).
When $\mathbf{h}_l \in \mathcal{W}^{\perp}(n)$, the corresponding $R_l$ is always zero
which is the case (i).

In cases 2) and 4), there exists a vector $\mathbf{h} \in C(n,f^{\perp}) \cap \mathcal{W}^{\perp}(n)$,
where $\cap$ denotes the intersection.
When $C(n,f^{\perp}) \cap \mathcal{W}^{\perp}(n)= \mathbf{0}_n$ we have $\mathbf{h} = \mathbf{0}_n$,
otherwise there exists a vector $\mathbf{h} \neq \mathbf{0}_n$ that belongs to the intersection space $C(n,f^{\perp}) \cap \mathcal{W}^{\perp}(n)$.
Let us first consider the case when there exists a vector $\mathbf{h} \in C(n,f^{\perp}) \cap \mathcal{W}^{\perp}(n)$
such that $\mathbf{h} \neq \mathbf{0}_n$. From Property~\ref{Property_division2}, the
vector space spanned by $\{ \mathbf{h}_0, \mathbf{h}_1, \ldots, \mathbf{h}_{\deg(f)-1} \}$ is equal to the code $C(n,f^{\perp})$ and
hence $\mathbf{h} \in C(n,f^{\perp})$ can be written as
\begin{align}
\mathbf{h} = a_0 \mathbf{h}_0 + a_1 \mathbf{h}_1 + \ldots + a_{\deg(f)-1} \mathbf{h}_{\deg(f)-1},
\label{Eqn_ai_lin_combi}
\end{align}
where each $a_l \in \mathbb{F}_2$ for $l = 0,1,\ldots, \deg(f)-1$ such that for some $i$, $0 \leq i < \deg(f)$,
$a_i \neq 0$. 
Since $\mathbf{h} \in \mathcal{W}^{\perp}(n)$, we have $\mathbf{w} \mathbf{h}^T = 0$ and
from (\ref{Eqn_ai_lin_combi}) we get
\begin{equation}
\begin{aligned}
\mathbf{w} \Big (a_0 \mathbf{h}_0 + a_1 \mathbf{h}_1 + \ldots + a_{\deg(f)-1} \mathbf{h}_{\deg(f)-1} \Big)^T &= 0 \\ 
\implies a_0 \mathbf{w} \mathbf{h}_0^T + a_1 \mathbf{w} \mathbf{h}_1^T + \ldots + a_{\deg(f)-1} \mathbf{w} \mathbf{h}_{\deg(f)-1}^T &= 0 \\ 
\implies a_0 r_0 + a_1 r_1 + \ldots + a_{\deg(f)-1} r_{\deg(f)-1} &= 0,
\label{Eqn_r_i_lin_combi_next}
\end{aligned}
\end{equation}
where the last equality is obtained from (\ref{Eqn_r_X}). Observe that this corresponds to the case (ii) 
when the set of random variables $\{R_0, R_1, \ldots, R_{\deg(f)-1}\}$ satisfy a linear relation.

We next consider the case when only the all-zero vector exists in the intersection of $C(n,f^{\perp})$ and 
$\mathcal{W}^{\perp}(n)$, i.e., in (\ref{Eqn_ai_lin_combi}), $\mathbf{h} = \mathbf{0}_n$.
From (\ref{Eqn_ai_lin_combi}) and (\ref{Eqn_r_i_lin_combi_next}) this implies that, the set of 
random variables $\{ R_0, R_1, \ldots, R_{\deg(f)-1} \}$ do not satisfy any linear relation.
Thus the set of random variables $\{ R_0, R_1, \ldots, R_{\deg(f)-1} \}$ are independent.
We now prove by contradiction that each $R_l$ is equally likely to be zero or one.
Suppose for some $i$, $0 \leq i < \deg(f)$, $R_i$ is always zero, which
implies that $\mathbf{h}_i \in \mathcal{W}^{\perp}(n)$.
Since $\mathbf{h}_i \in C(n,f^{\perp})$ we have $\mathbf{h}_i \in \mathcal{W}^{\perp}(n) \cap C(n,f^{\perp})$
such that $\mathbf{h}_i \neq \mathbf{0}_n$, which is a contradiction.
Note that this situation corresponds to case (iii) and the proof is complete.

We now consider the case when $\mathcal{W}(n)$ is a trivial code. 
When $\mathcal{W}(n)$ contains only the all-zero codeword, $r(X)$ will be zero with probability one and follows 
the degenerate distribution.
When $\mathcal{W}(n) = \mathbb{F}_2^n$, since $\mathbf{w}(X)$ takes any value in $\mathcal{W}(n)$
with the uniform distribution, the random variable corresponding to $r(X) = \mathbf{w}(X) \bmod f(X)$ 
follows the uniform distribution and the proof is complete.
\end{IEEEproof}
We now use this lemma to prove Proposition~\ref{Proposition_cyclic_structure_no_type1}.

~ \\~
\textbf{\textit{Proof of Proposition~\ref{Proposition_cyclic_structure_no_type1}:}}\\
Recall that the subspace $\mathcal{W}(n)$ is obtained by considering the initial $n$ bits of codewords of $C(n_0,g_0)$
and $\mathcal{W}^{\prime}(n)$ is defined in (\ref{Eqn_Wn_prime_pre_suff_definition}).
In \cite{TCOMM_2016}, Yardi et al.~proved that 
there exists a codeword $\mathbf{w}_1(X) \in \mathcal{W}(n)$ and a codeword $\mathbf{w}_1^{\prime}(X) \in \mathcal{W}^{\prime}(n)$
such that $\mathbf{w}_1(X), \mathbf{w}_1^{\prime}(X) \notin C(n,f)$, where $C(n,f)$ is the cyclic code generated by $f(X)$
(see Appendix~B, Proposition~1 of~\cite{TCOMM_2016}).
For $\mathbf{w}_1(X)$ and $\mathbf{w}_1^{\prime}(X)$, the corresponding syndromes $r(X) = \mathbf{w}_1(X) \bmod f(X)$ 
and $r^{\prime}(X) = \mathbf{w}_1^{\prime}(X) \bmod f(X)$ will be nonzero polynomials.
Since the all-zero vector is always a codeword in any linear block code, $r(X)$ and $r^{\prime}(X)$ will be the zero polynomial
for the all-zero codeword. 
Since $r(X)$ and $r^{\prime}(X)$ can take at least two values in $\mathcal{P}_{\deg(f)}$ with a non-zero probability, 
$r(X)$ and $r^{\prime}(X)$ cannot follow the degenerate distribution (see (\ref{Eqn_type_1_distri})).
From Lemma~\ref{Lemma_three_type_distri}, the distribution of the random variables corresponding to
$r(X)$ and $r^{\prime}(X)$ will either be uniform or restricted uniform and the proof is complete.
\hfill $\blacksquare$

\section*{Appendix C: Proof of Theorem~\ref{Theorem_main_iff_structure}}
%
We first prove the necessary condition of the theorem that, if $r(X) = \mathbf{w}(X) \bmod f(X)$ follows the restricted uniform distribution,
$g_0^{\perp}(X)$ has a factor of order strictly less than $n_0$.
Let $\mathcal{W}(n)$ be the vector space obtained by puncturing the last $n_0-n$ bits of codewords in code $C(n_0,g_0)$
such that $\mathbf{w}(X) \in \mathcal{W}(n)$.
Since $C(n_0,g_0)$ is a cyclic code, the initial $k_0$ bits can be considered an information set and 
the assumption $k_0 < n < n_0$ implies that the dimension of $\mathcal{W}(n)$ is $k_0$.

Suppose $r(X) = \mathbf{w}(X) \bmod f(X)$ is given by,
\begin{equation}
\begin{aligned}
r(X) &= r_0 +r_1X + \ldots + r_{{\deg(f)}-1}X^{{\deg(f)}-1} \\ 
	      &= \mathbf{w}\mathbf{h}_0^T +\mathbf{w}\mathbf{h}_1^T X + \ldots + \mathbf{w}\mathbf{h}_{{\deg(f)}-1}^TX^{{\deg(f)}-1},
\label{Eqn_r_w_mod_f_proof}
\end{aligned}
\end{equation}
where the last equality is obtained from Property~\ref{Property_division1} of Appendix~B such that each $\mathbf{h}_l \in \mathbb{F}_2^n$
for $l = 0,1,\ldots,\deg(f)-1$.
From Property~\ref{Property_division2} of Appendix~B, each $\mathbf{h}_l \in C(n,f^{\perp})$
where $C(n,f^{\perp})$ is the dual code of the cyclic code $C(n,f)$
and the vector space spanned by $\{ \mathbf{h}_0, \mathbf{h}_1, \ldots, \mathbf{h}_{\deg(f)-1} \}$ is equal to $C(n,f^{\perp})$.

As explained in the proof of Lemma~\ref{Lemma_three_type_distri}, the random variable corresponding to 
$r(X)$ follows the restricted uniform distribution if and only if 
there exists a non-zero vector $\mathbf{h} \in \mathbb{F}_2^n$ that lies in the intersection
space of the codes $C(n,f^{\perp})$ and $\mathcal{W}^{\perp}(n)$ (see Appendix~B).
Since $\mathbf{h} \in C(n,f^{\perp})$, for some $\mathbf{u}_1(X) \in \mathcal{P}_{n-\deg(f^{\perp})}$ we have
\begin{align}
\mathbf{h}(X) = \mathbf{u}_1(X) f^{\perp}(X).
\label{Eqn_in_f_perp}
\end{align}
We now prove that the vector $\mathbf{h}^{\prime} \coloneqq [\mathbf{h} \mbox{~~} \mathbf{0}_{n_0-n}]$ lies in the code $C(n_0, g_0^{\perp})$.
For any $\mathbf{v} \in C(n_0,g_0)$, the inner product of $\mathbf{v}$ and $\mathbf{h}^{\prime}$ is given by
\begin{equation}
\begin{aligned}
\mathbf{v} (\mathbf{h}^{\prime})^T &= \Big[\mathbf{v}(0:n-1) \mbox{~~}\mathbf{v}(n:n_0-1)\Big] \Big[\mathbf{h} \mbox{~~} \mathbf{0}_{n_0-n}\Big]^T \\
				   &= \mathbf{v}(0:n-1) \mathbf{h}^T \\
				   &= \mathbf{w} \mathbf{h}^T  \\
				   &= 0, 
\end{aligned}
\label{Equation_W_perp_in_Cn0g0_perp}
\end{equation}
where the last equality is obtained since $\mathbf{h} \in \mathcal{W}^{\perp}(n)$.
Since $\mathbf{h}^{\prime} \in C(n_0,g_0^{\perp})$, for some $\mathbf{u}_2(X) \in \mathcal{P}_{n_0-k_0}$ we have
\begin{align}
\mathbf{h}^{\prime}(X)= \mathbf{h}(X) = \mathbf{u}_2(X) g_0^{\perp}(X).
\label{Eqn_in_g0_perp}
\end{align}
Equating (\ref{Eqn_in_f_perp}) and (\ref{Eqn_in_g0_perp}) we get
\begin{align}
\mathbf{h}(X) = \mathbf{u}_1(X) f^{\perp}(X) = \mathbf{u}_2(X) g_0^{\perp}(X).
\label{Eqn_degree_inequality}
\end{align}

From the assumption of the theorem, $\deg(f) \leq k_0$ which implies that $n-\deg(f) \geq n-k_0$.
Since $\deg(f^{\perp}) = n - \deg(f)$, we have $\deg(f^{\perp}) \geq n-k_0$. Since $\deg(g_0^{\perp}) = k_0$
and $\deg(h) \leq n-1$, from (\ref{Eqn_degree_inequality}) we get $\deg(u_2) \leq n-k_0-1$.
Thus in (\ref{Eqn_degree_inequality}) we have $\deg(u_2) \leq n-k_0-1$ and $\deg(f^{\perp}) \geq n-k_0$.
This implies that there exists a factor $f_1(X)$ of $f^{\perp}(X)$ such that $f_1(X)$ is a factor of $g_0^{\perp}(X)$.
Since $f_1(X)$ is a factor of $X^n+1$ and $n < n_0$, this implies that $g_0^{\perp}(X)$ has a factor of order
strictly less than $n_0$ and
the proof of the necessary condition is complete.


We will now prove the converse. 
Suppose $g_0^{\perp}(X)$ has a factor $m^{\perp}(X)$ of order $n^{\prime}$ such that $1 \leq n^{\prime} < n$.
For a non-degenerate code $C(n_0,g_0)$, the order of $g_0^{\perp}(X)$ is equal to $n_0$~\cite[Sec.~8.3]{Macwilliams_Sloane_1977} 
and hence $m^{\perp}(X) \neq g_0^{\perp}(X)$.
From Lemma~\ref{Lemma_dual_gen_factor_degenerate_pattern} of Appendix~A, this implies that there exists a 
codeword of a degenerate pattern in $C(n_0,g_0)$, i.e., there exists
$\mathbf{v} \in C(n_0,g_0)$ given by,
\begin{align}
\mathbf{v} = \Big[\underbrace{\mathbf{w}^{\prime} \mbox{~~} \mathbf{w}^{\prime} \mbox{~~} \cdots \mbox{~~} \mathbf{w}^{\prime}}_{l \text{ times}}\Big],
\label{Eqn_v_app_LRR2}
\end{align}
where $l > 1$, $\mathbf{w}^{\prime}$ is vector of length $n^{\prime}$ such that $\mathbf{w}^{\prime}$ is not a vector of 
a degenerate pattern (see Definition~\ref{Definition_degenerate_pattern}).
Note that $m^{\perp}(X)$ is the minimal polynomial polynomial of the linear recurring sequence
given by $[\mathbf{w}^{\prime} \mbox{~} \mathbf{w}^{\prime} \mbox{~} \cdots]$~\cite[Sec.~8.3]{Macwilliams_Sloane_1977}. 
It is given that, $m(X)$ is the minimal generating polynomial of this sequence.
Thus each $\mathbf{w}^{\prime}(X)$ is a multiple of $m(X)$ (see Definition~\ref{Definition_min_gen_poly_LRR}).
Suppose $\mathbf{w}^{\prime}(X) = \mathbf{u}^{\prime}(X) m(X)$, for some $\mathbf{u}^{\prime}(X) \in \mathcal{P}_{\deg(m^{\perp})}$,
since $\deg(m^{\perp}) = n^{\prime}-\deg(m)$.
Substituting this in (\ref{Eqn_v_app_LRR2}) we get,
\begin{align}
\mathbf{v}(X) &= \mathbf{w}^{\prime}(X) + X^{n^{\prime}}\mathbf{w}^{\prime}(X) + \ldots + X^{(l-1)n^{\prime}}\mathbf{w}^{\prime}(X) \\
	      &= \mathbf{u}^{\prime}(X) m(X) + X^{n^{\prime}}\mathbf{u}^{\prime}(X) m(X) + \ldots + \nonumber \\
	      &\mbox{~~~~~~~~~~~~~~~~~~} X^{(l-1)n^{\prime}}\mathbf{u}^{\prime}(X) m(X) \\	    
	      &= \mathbf{u}^{\prime}(X) m(X) \Big( 1+ X^{n^{\prime}} + \ldots + X^{(l-1)n^{\prime}} \Big)
\label{Eqn_num_w_multi_m0}	      
\end{align}

Let $C^{\perp}(n_0,m^{\perp})$ be the dual code of $C(n_0,m^{\perp})$, 
where $C(n_0,m^{\perp})$ is the cyclic code of length $n_0$ generated by $m^{\perp}(X)$. 
Note that $\mathbf{v}(X) \in C^{\perp}(n_0,m^{\perp})$ and from (\ref{Eqn_num_w_multi_m0}), 
the set of codewords in $C^{\perp}(n_0,m^{\perp})$ are obtained considering all possible $2^{\deg(m^{\perp})}$ 
values of $\mathbf{u}^{\prime}(X) \in \mathcal{P}_{\deg(m^{\perp})}$.
Since $m^{\perp}(X)$ is a factor of $g_0^{\perp}(X)$ we have 
$C(n_0,g_0^{\perp}) \subset C(n_0,m^{\perp})$ and this implies that
$C^{\perp}(n_0,m^{\perp}) \subset C(n_0,g_0)$.
Thus the codewords in $C(n_0,g_0)$ that are multiples of $m(X)$ are exactly the $2^{\deg(m^{\perp})}$ codewords
in $C^{\perp}(n_0,m^{\perp})$. 


From the assumptions of the converse, we have $n = bn^{\prime}$ for some $b \geq 1$.
Thus the vector $\mathbf{w}$ formed by the initial $n$ bits of $\mathbf{v}$ in (\ref{Eqn_v_app_LRR2}) is given by,
\begin{align}
\mathbf{w} = \Big[\underbrace{\mathbf{w}^{\prime} \mbox{~~} \mathbf{w}^{\prime} \mbox{~~} \cdots \mbox{~~} \mathbf{w}^{\prime}}_{b \text{ times}}\Big].
\label{Eqn_w_app_LRR11}
\end{align}
Substituting $\mathbf{w}^{\prime}(X) = \mathbf{u}^{\prime}(X) m(X)$ we get, 
\begin{align}
\mathbf{w}(X) &= \mathbf{w}^{\prime}(X) + X^{n^{\prime}}\mathbf{w}^{\prime}(X) + \ldots + X^{(b-1)n^{\prime}}\mathbf{w}^{\prime}(X) \nonumber \\
	      &= \mathbf{u}^{\prime}(X) m(X) \Big( 1+ X^{n^{\prime}} + \ldots + X^{(b-1)n^{\prime}} \Big).
\label{Eqn_num_w_multi_m}	      
\end{align}

As explained in the first paragraph of the proof, the dimension of $\mathcal{W}(n)$ is $k_0$ and hence
corresponding to every $\mathbf{v} \in C(n_0,g_0)$ there is a unique $\mathbf{w} \in \mathcal{W}(n)$.
From (\ref{Eqn_num_w_multi_m0}) and (\ref{Eqn_num_w_multi_m}), 
this implies that the number of $\mathbf{w}(X) \in \mathcal{W}(n)$ that are multiples of $m(X)$ 
are equal to $2^{\deg(m^{\perp})}$.
From (\ref{Eqn_num_w_multi_m}), any $\mathbf{w}(X) \in \mathcal{W}(n)$ that is a multiple of $m(X)$ is also a
multiple of $(1+ X^{n^{\prime}} + \ldots + X^{(b-1)n^{\prime}})$. 
For a factor $f(X)$ of $m(X)(1+X^{n^{\prime}}+X^{2n^{\prime}}+\ldots+X^{(b-1)n^{\prime}})$, 
the probability that $r(X) = \mathbf{w}(X) \bmod f(X)$ is the all-zero polynomial is given by,
\begin{align}
&\mathbb{P} \big[r(X) = 0\big] = \mathbb{P} \big[ \mathbf{w}(X) \bmod f(X) = 0 \big] \nonumber \\
		      &= \frac{\mbox{Number of }\mathbf{w}(X) \in \mathcal{W}(n) \mbox{ that are multiples of } f(X)}{ \mbox{Total number of } \mathbf{w}(X) \in \mathcal{W}(n)}\nonumber  \\
		      &= \frac{2^{\deg(m^{\perp})}}{2^{k_0}}\nonumber  \\
		      & \stackrel{(a)}{>}  \frac{2^{k_0 - \deg(f)}}{2^{k_0}} \nonumber \\
		      &= \frac{1}{2^{\deg(f)}}
\label{Eqn_num_w_multi_m_new}			      
\end{align}
where the inequality in $(a)$ is obtained since $\deg(m^{\perp}) > k_0-\deg(f)$.

From Proposition~\ref{Proposition_cyclic_structure_no_type1},
the random variable corresponding to $r(X)$ can either follow the uniform distribution or the restricted uniform distribution.
%
For the uniform distribution, the probability of zero syndrome is equal to $1/2^{\deg(f)}$.
From (\ref{Eqn_num_w_multi_m_new}), the probability of zero syndrome is strictly more than $1/2^{\deg(f)}$
and hence $r(X)$ should follow the restricted uniform distribution. This completes the proof of the converse.
\hfill $\blacksquare$


\section*{Appendix D: Proof of Theorem~\ref{Theorem_W_suff_pre_cyclic_structure}}
%
Recall that the subspace $\mathcal{W}(n)$ is obtained by considering the initial $n$ bits of codewords
of $C(n_0,g_0)$. Since $ n< n_0$ from (\ref{Eqn_Wn_prime_pre_suff_definition}) we have 
$\mathcal{W}^{\prime}(n) = C_1(d_1) + C_2(d_2)$, where 
$C_1(d_1)$ and $C_2(d_2)$ are the linear block codes obtained by 
considering the set of suffixes and prefixes of lengths $d_1$ and $d_2$ of codewords in $C(n_0,g_0)$ respectively.
Note that due to the cyclic nature, the subspaces spanned by the set of prefixes of length $d_1$
and the set of suffixes of length $d_1$ are identical.
This implies that the code $\mathcal{W}^{\prime}(n)$ consists of all possible prefixes of length $d_1$ concatenated with
all possible suffixes of length $d_2$ and hence,
\begin{align}
\mathcal{W}(n) \subseteq \mathcal{W}^{\prime}(n).
\label{Eqn_W_subset_W_prime}
\end{align}
From (\ref{Eqn_W_subset_W_prime}) we have,
\begin{align}
\mathcal{W}^{\prime \perp}(n) \subseteq \mathcal{W}^{\perp}(n),
\label{Eqn_W_prime_perp_subset_W_perp}
\end{align}
where $\mathcal{W}^{\prime \perp}(n)$ and $\mathcal{W}^{\perp}(n)$ are the dual codes of 
$\mathcal{W}^{\prime}(n)$ and $\mathcal{W}(n)$ respectively.

In order to prove that $r^{\prime}(X) = \mathbf{w}^{\prime}(X) \bmod f(X)$ for $\mathbf{w}^{\prime}(X) \in \mathcal{W}^{\prime}(n)$
follows the uniform distribution using the arguments similar to the proof of Proposition~\ref{Proposition_cyclic_structure_no_type1},
we need to prove that $\mathcal{W}^{\prime \perp}(n) \cap C(n,f^{\perp}) = \mathbf{0}_n$.
From the assumptions of the theorem, $r(X) = \mathbf{w}(X) \bmod f(X)$ for $\mathbf{w}(X) \in \mathcal{W}(n)$
follows the uniform distribution. 
Using the arguments similar to the proof of Proposition~\ref{Proposition_cyclic_structure_no_type1},
this is possible when $\mathcal{W}^{\perp}(n) \cap C(n,f^{\perp}) = \mathbf{0}_n$.
From (\ref{Eqn_W_prime_perp_subset_W_perp}), this implies that $\mathcal{W}^{\prime \perp}(n) \cap C(n,f^{\perp}) = \mathbf{0}_n$
and the proof is complete.
\hfill $\blacksquare$


\section*{Appendix E: Proof of Theorem~\ref{Theorem_structure_w_prime_type3}}
%

Let us first consider that case when $q=0$, i.e., $r^{\prime}(X)$ is given by,
\begin{align}
r^{\prime}(X) = t_1(X) + t_2(X).
\label{Eqn_t1_t2_2}	      
\end{align}
We now consider the situation when both $t_1(X)$ and $t_2(X)$ follow the uniform distribution.
The probability that $r^{\prime}(X)$ is a zero polynomial is given by,
\begin{eqnarray}
\begin{aligned}
&\mathbb{P} \big[r^{\prime}(X) = 0\big] 
\stackrel{(a)}{=} \mathbb{P} \big[ t_1(X) + t_2(X) = 0\big] \\
& = \mathbb{P} \big[ t_1(X) = t_2(X)\big] \\
& = \sum_{a(X) \in \mathcal{P}_{\deg(f)}} \mathbb{P} \Big[ t_1(X) = t_2(X) = a(X)\Big]\\
& \stackrel{(b)}{=} \sum_{a(X) \in \mathcal{P}_{\deg(f)}} \mathbb{P} \Big[ t_1(X) = a(X)\Big] \mathbb{P} \Big[t_2(X) = a(X)\Big]\\
& \stackrel{(c)}{=} \sum_{a(X) \in \mathcal{P}_{\deg(f)}} \frac{1}{2^{\deg(f)}} \frac{1}{2^{\deg(f)}} 
= \frac{1}{2^{\deg(f)}}.
\label{Eqn_r_prime_equally}
\end{aligned}
\end{eqnarray}
The equality in $(a)$ is obtained from (\ref{Eqn_t1_t2_2}) and $(b)$, $(c)$ follow since the random
variables corresponding to $t_1(X)$ and $t_2(X)$ are i.i.d.~according
to the uniform distribution.
From Proposition~\ref{Proposition_cyclic_structure_no_type1},
the random variable corresponding to $r^{\prime}(X)$ can either follow the uniform distribution or the restricted uniform distribution. 
From (\ref{Eqn_r_prime_equally}), the random variable corresponding to 
$r^{\prime}(X)$ follows the uniform distribution.

We next consider the case when either $t_1(X)$ or $t_2(X)$ follow the restricted uniform distribution.
Without loss of generality let us consider the case when
$t_1(X)$ follows the restricted uniform distribution. 
From the definition of the restricted uniform distribution we get, $\mathbb{P}[ t_1(X) = a(X)] > 1/2^{\deg(f)}$ 
and in (\ref{Eqn_r_prime_equally}) we have
\begin{align}
\mathbb{P} \Big[r^{\prime}(X) = 0\Big] > \frac{1}{2^{\deg(f)}}.
\label{Eqn_num_w_multi_m_new22}
\end{align}
As explained earlier, the random variable corresponding to $r^{\prime}(X)$ can either follow the uniform distribution 
or the restricted uniform distribution.
For the uniform distribution, the probability of zero syndrome should be equal to $1/2^{\deg(f)}$.
From (\ref{Eqn_num_w_multi_m_new22}), the probability of zero syndrome is more than $1/2^{\deg(f)}$
and hence $r^{\prime}(X)$ follows the restricted uniform distribution. This completes the proof for the case when $q=0$.

The case when $q>0$ can be proved using similar arguments and hence we will not discuss it in detail.
\hfill $\square$

\section*{Appendix F: Proof of Theorem~\ref{Theorem_lower_bound_H1_PCnf}}
%
Since the proof is the same for any $j$th received polynomial $\mathbf{y}_j(X)$, for simplicity of notation we will ignore the suffix $j$ from 
$\mathbf{y}_j(X)$ in this proof. Using this, the received polynomial $\mathbf{y}(X)$ is given by,

\begin{align}
\mathbf{y}(X) = \mathbf{w}(X) + \mathbf{e}(X),
\label{Eqn_y_w_e}
\end{align}
where $\mathbf{w}(X)$ is the error-free polynomial and $\mathbf{e}(X)$ is the polynomial corresponding to the error introduced
by BSC($p$). The probability of observing the all-zero syndrome is given by,
\begin{equation}
\begin{aligned}
\mathbb{P}\Big[r(X) = 0\Big] &= \mathbb{P}\Big[\mathbf{y}(X) \bmod f(X) = 0\Big] \\
&= \mathbb{P}\Big[\mathbf{y}(X) \in C(n,f)\Big]
\end{aligned}
\label{Eqn_long1}
\end{equation}
where the last equality is obtained since the cyclic code $C(n,f)$ consists of possible
multiples of $f(X)$. 
For a given $\mathbf{w}(X)$ there are two possibilities, either 
$\mathbf{w}(X) \in C(n,f)$ or $\mathbf{w}(X) \notin C(n,f)$.
Suppose,
\begin{equation}
\begin{aligned}
Q &: \mbox{ Event when } \mathbf{w}(X) \in C(n,f), \\
Q^c &: \mbox{ Event when } \mathbf{w}(X) \notin C(n,f).
\end{aligned}
\label{Eqn_Q_Qc_event}
\end{equation}
Using total probability law in (\ref{Eqn_long1}) we get,
\begin{align}
\mathbb{P}\Big[r(X) = 0\Big] & = \mathbb{P}\Big[\mathbf{y}(X) \in C(n,f) \Big| Q \Big] \mathbb{P}[Q] + \nonumber \\
&\mbox{~~~~~~~~~~~~}\mathbb{P}\Big[\mathbf{y}(X) \in C(n,f) \Big| Q^c \Big] \mathbb{P}[Q^c].
\label{Eqn_long2}
\end{align}

From (\ref{Eqn_y_w_e}) and (\ref{Eqn_Q_Qc_event}), when the event $Q$ is true, we have $\mathbf{y}(X) \in C(n,f)$ if $\mathbf{e}(X) \in C(n,f)$.
Similarly, when the event $Q^c$ is true, we have $\mathbf{y}(X) \in C(n,f)$ if $\mathbf{e}(X)$
belongs to some proper coset $\mathcal{G}(n,f)$ of code $C(n,f)$.
Using this in (\ref{Eqn_long2}) we have,
\begin{align}
\mathbb{P}\Big[r(X) = 0\Big] &= \mathbb{P} \Big[\mathbf{e}(X) \in C(n,f) \Big] \mathbb{P}[Q] + \nonumber \\
&\mbox{~~~~~~~~~~~~} \mathbb{P} \Big[ \mathbf{e}(X) \in \mathcal{G}(n,f)) \Big] \mathbb{P}[Q^c]
\label{Eqn_long3}
\end{align}

From Sullivan's subgroup-coset inequality theorem~\cite{Sullivan67}, 
for any proper coset $\mathcal{G}(n,f)$ of $C(n,f)$ we have,
\begin{align}
\frac{\mathbb{P}[\mathbf{e}(X) \in C(n,f)]}{\mathbb{P}[\mathbf{e}(X) \in \mathcal{G}(n,f)]} 
\geq \frac{1-(1-2p)^{n-\deg(f)+1}}{1+(1-2p)^{n-\deg(f)+1}} = \lambda.
\label{Eqn_sullivan_app_structure}
\end{align}
We next find the probability of the event $\mathbf{e}(X) \in C(n,f)$ as follows.
\begin{equation}
\begin{aligned}
\mathbb{P} [\mathbf{e}(X) \in C(n,f)] &= \sum_{\mathbf{v}(X) \in C(n,f)}  \mathbb{P} [\mathbf{e}(X) = \mathbf{v}(X)] \\
		    &= \sum_{i=0}^n A_i p^i (1-p)^{n-i} \\
		    & = P(C(n,f)),
\end{aligned}
\label{Eqn_PCf_definition_repeat}
\end{equation}
where $\{ A_0, A_1, \cdots, A_n\}$ is the weight distribution of $C(n,f)$ and last equality is obtained from (\ref{Eqn_PCf_definition}).
Substituting (\ref{Eqn_sullivan_app_structure}) in (\ref{Eqn_long3}) we have,
\begin{align}
\mathbb{P}\Big[r(X) &= 0\Big] 
 \leq \mathbb{P} \Big[\mathbf{e}(X) \in C(n,f) \Big] \mathbb{P}[Q] + \nonumber \\
& \mbox{~~~~~~~~~~~~~~~~~~} \lambda \mathbb{P}\Big[ \mathbf{e}(X) \in C(n,f)) \Big] \mathbb{P}[Q^c] \\
& \stackrel{(b)}{=} \mathcal{P}(C(n,f)) \mathbb{P}[Q] + \lambda \mathcal{P}(C(n,f)) \big(1-\mathbb{P}[Q]\big)\\
& = \mathcal{P}(C(n,f)) \Big[ \mathbb{P}[Q] + \lambda \big(1-\mathbb{P}[Q]\big) \Big]\\
& = \mathcal{P}(C(n,f)) \Big[ \mathbb{P}[Q] (1 - \lambda) + \lambda \Big]
%
%
\label{Eqn_long4}
\end{align}
The equality in $(b)$ is obtained from (\ref{Eqn_PCf_definition_repeat}) and since $\mathbb{P}[Q^c] = 1 - \mathbb{P}[Q]$ (see (\ref{Eqn_Q_Qc_event})).

From the assumption of the theorem, either $n \neq ln_0$ or assumed synchronization $s \neq s_0$ or
$f(X)$ is not a factor of $g_0(X)$.
When either $n \neq ln_0$ or $s \neq s_0$ or $f(X)$ is not a factor of $g_0(X)$, from Proposition~\ref{Proposition_cyclic_structure_no_type1} 
and Section~\ref{subsection_n_ln0_s_s0}, the distribution of $\mathbf{w}(X) \bmod f(X)$ is either uniform or restricted uniform.
From the definition of the uniform and the restricted uniform distributions,
$\mathbb{P}[\mathbf{w}(X) \bmod f(X) = 0]$ is less than or equal to $1/2$, i.e., 
\begin{align}
\mathbb{P}[\mathbf{w}(X) \in C(n,f)]  = \mathbb{P}[Q] \leq \frac{1}{2}.
\label{Eqn_PQ_less_1_2}
\end{align}
Substituting (\ref{Eqn_PQ_less_1_2}) in (\ref{Eqn_long4}) we get,
\begin{align}
\mathbb{P}\Big[r(X) = 0\Big] 
& \leq \mathcal{P}(C(n,f)) \left( \frac{1}{2} (1 - \lambda) + \lambda \right)\\
& = \mathcal{P}(C(n,f)) \left( \frac{\lambda+1}{2} \right)
\label{Eqn_long5}
\end{align}
and the proof is complete.
\hfill $\blacksquare$


\section*{Appendix G: Proof of Theorem~\ref{Theorem_chinese}}
%
Since the proof is the same for any $j$th received vector $\mathbf{y}_j$, we will ignore the suffix $j$
from $\mathbf{y}_j$ for the sake of simplicity.
Using this notation, an $n$-bit received vector is given by,
\begin{align}
\mathbf{y} = \mathbf{w} + \mathbf{e},
\label{Eqn_chinese_y_w_e}
\end{align}
where $\mathbf{w}$ is an error-free vector and $\mathbf{e}$ is an error vector introduced by BSC($p$).
For a factor $f(X)$ of $X^n+1$, suppose a parity check matrix $H$ of $C(n,f)$ is given by
\begin{align}
H & = \begin{bmatrix}
      f^{\perp}_0    & f^{\perp}_{1}   & \cdot  & \cdot   & f^{\perp}_{\deg(f^{\perp})} & 0 & \cdot & 0 \\
      0      & f^{\perp}_0   & \cdot  & \cdot   & \cdot   & f^{\perp}_{\deg(f^{\perp})}   & \cdot & 0 \\
      \vdots &  &  \ddots  &  &         &         & \vdots    &   \\
      0      & \cdot & 0      & f^{\perp}_0     & \cdot   & \cdot     & \cdot & f^{\perp}_{\deg(f^{\perp})}
    \end{bmatrix} \nonumber \\
    & =
    \begin{bmatrix}
      \mathbf{h}_0 \\
      \mathbf{h}_1 \\
      \vdots \\
      \mathbf{h}_{\deg(f)-1}
    \end{bmatrix},
\label{Eqn_H_Cnf_app}    
\end{align}
where the polynomial corresponding to the first row of $H$ is the generator polynomial $f^{\perp}(X)$ of the dual code of $C(n,f)$, and 
$\mathbf{h}_0, \mathbf{h}_1, \ldots, \mathbf{h}_{\deg(f)-1}$ are the rows of $H$.
Suppose $\mathbf{w}H^T$ is given by,
\begin{align}
\mathbf{w} H^T = \mathbf{t} &= \begin{bmatrix} \mathbf{w}\mathbf{h}_1^T & \mathbf{w}\mathbf{h}_2^T & \ldots & \mathbf{w}\mathbf{h}_{\deg(f)-1}^T \end{bmatrix}\\
			   &= \begin{bmatrix} t_0 & t_1 & \ldots & t_{\deg(f)-1}\end{bmatrix} 			   
\label{Eqn_tl}			   
\end{align}
where $t_l = \mathbf{w}\mathbf{h}_l^T$, for $l = 0,1,\ldots,\deg(f)-1$.
As shown in \Fig~\ref{Figure_n_more_less_n0_noise_free}, an $n$-bit noise-free vector $\mathbf{w}$ is either 
of the following two types.
\begin{enumerate}[(i)]
\item $\mathbf{w}$ is formed by the consecutive $n$ bits of a codeword in the true code $C(n_0,g_0)$, i.e., 
$\mathbf{w} \in \mathcal{W}(n)$, where $\mathcal{W}(n)$ is defined in the first paragraph of Section~\ref{Section_cyclic_structure_noise_free}.
\item $\mathbf{w}$ is a concatenation of the suffix of a codeword of length $d_1$, a sequence of $q$ codewords,
and the prefix of a codeword of length $d_2$, where $0 \leq d_1,d_2 < n_0$, $q \geq 1$ such that $n=d_1+qn_0+d_2$, i.e., 
$\mathbf{w} \in \mathcal{W}^{\prime}(n)$, where $\mathcal{W}^{\prime}(n)$ is defined in (\ref{Eqn_Wn_prime_pre_suff_definition}).
\end{enumerate}
We now consider the cases when $\mathbf{w} \in \mathcal{W}(n)$ and $\mathbf{w} \in \mathcal{W}^{\prime}(n)$ 
separately and prove that $t_l$ in (\ref{Eqn_tl}) is equally likely to be zero or one for $l = 0,1,\ldots, \deg(f)-1$.
\begin{enumerate}[(i)]
\item Case when $\mathbf{w}\in \mathcal{W}(n)$ \\
From the assumptions of the theorem we have $n < n_0$.
For a given $\mathbf{h}_l$ we have either $\mathbf{h}_l \in \mathcal{W}^{\perp}(n)$ or
$\mathbf{h}_l \notin \mathcal{W}^{\perp}(n)$, where $\mathcal{W}^{\perp}(n)$ is the dual code of $\mathcal{W}(n)$. 
We now prove by contradiction that each $\mathbf{h}_l \notin \mathcal{W}^{\perp}(n)$, for $l = 0,1,\ldots, \deg(f)-1$.
Suppose $\mathbf{h}_l \in \mathcal{W}^{\perp}(n)$ for some $l$, $0 \leq l < \deg(f)$.
Using the similar steps as in (\ref{Equation_W_perp_in_Cn0g0_perp}) we can prove that, 
$[\mathbf{h}_l \mbox{~~} \mathbf{0}_{n_0-n}] \in C(n_0,g_0^{\perp})$, where $C(n_0,g_0^{\perp})$ is the dual code of 
$C(n_0,g_0)$. From (\ref{Eqn_H_Cnf_app}), 
the polynomial corresponding to $\mathbf{h}_l$ can be written as $\mathbf{h}_l(X) =  X^l f^{\perp}(X)$
and $[\mathbf{h}_l \mbox{~~} \mathbf{0}_{n_0-n}] \in C(n_0,g_0^{\perp})$ implies that,
\begin{align}
\mathbf{h}_l(X) = X^l f^{\perp}(X) = \mathbf{u}(X)g_0^{\perp}(X),
\label{Eqn_app_g0_perp_minimal}
\end{align}
where $\mathbf{u}(X) \in \mathcal{P}_{n_0-\deg(g_0^{\perp})}$.
For a nontrivial cyclic code, $g_0^{\perp}(X)$ does not divide $X^l$ for any integer $l$~\cite{LinCostello2004}, and 
hence (\ref{Eqn_app_g0_perp_minimal}) implies that $g_0^{\perp}(X)$ should divide $f^{\perp}(X)$.
Since $f^{\perp}(X)$ divides $X^n+1$, $g_0^{\perp}(X)$ also divides $X^n+1$.
Since $n < n_0$, $C(n_0,g_0)$ will be a degenerate code~\cite[Sec.~8.3]{Macwilliams_Sloane_1977}, 
which is a contradiction according to the assumptions of the theorem. 
This proves that $\mathbf{h}_l \notin \mathcal{W}^{\perp}(n)$ for $l = 0,1,\ldots, \deg(f)-1$.
From Lemma~\ref{Lemma_h_not_dual_equally}, $\mathbf{h}_l \notin \mathcal{W}^{\perp}(n)$ implies that $t_l$ 
is equally likely to be zero or one.

\item When $\mathbf{w} \in \mathcal{W}^{\prime}(n)$\\
Since $n < n_0$, an $n$-bit vector $\mathbf{w}$ is given by, 
\begin{align}
\mathbf{w} = \Big[ \mathbf{v}_{1}(n_0-d_1:n_0-1)  \mbox{~~} \mathbf{v}_{2}(0:d_2-1) \Big],
\end{align}
where $\mathbf{v}_1, \mathbf{v}_2 \in C(n_0,g_0)$.
For a given $\mathbf{h}_l$ the inner product $\mathbf{w}\mathbf{h}_l^T$ is given by,
\begin{align}
\mathbf{w} \mathbf{h}_l^T & = \mathbf{w}(0:d_1-1) \mathbf{h}_l(0:d_1-1)^T + \nonumber \\
&\mbox{~~~~~~~~~~~} \mathbf{w}(d_1:n-1) \mathbf{h}_l(d_1:n-1)^T.
\label{Eqn_wht}
\end{align}
Recall that in part (i) we proved that $\mathbf{h}_l \notin \mathcal{W}^{\perp}(n)$ for $l = 0,1,\ldots, \deg(f)-1$.
From Lemmas~\ref{Lemma_h_not_dual_equally} and \ref{Lemma_for_sync_equally_likely} of Appendix~A, $\mathbf{h}_l \notin \mathcal{W}^{\perp}(n)$ implies that 
either $\mathbf{w}(0:d_1-1) \mathbf{h}_l(0:d_1-1)^T$ or $\mathbf{w}(d_1:n-1) \mathbf{h}_l(d_1:n-1)^T$ is equally likely 
to be zero or one.
This implies that in (\ref{Eqn_wht}), $t_l = \mathbf{w} \mathbf{h}_l^T$ is equally likely to be zero or one.

%
\end{enumerate}

We now have that each bit in $\mathbf{w}\mathbf{h}_l$ is equally likely to be zero or one, for $l=0,1,\ldots,\deg(f)-1$.
Let us consider the noise-affected version of $\mathbf{y}$ of $\mathbf{w}$ (see (\ref{Eqn_chinese_y_w_e})).
Suppose the inner product $\mathbf{y}\mathbf{h}_l$ is given by,
\begin{align}
\mathbf{y} H^T = \mathbf{r} = \begin{bmatrix} r_0 & r_1 & \ldots & r_{\deg(f)-1}\end{bmatrix} 			     
\end{align}
where each $r_l$ is given by,
\begin{align}
r_l &= \mathbf{y} \mathbf{h}_l^T = \Big[\mathbf{w} + \mathbf{e} \Big] \mathbf{h}_l^T \\
    &= \mathbf{w}\mathbf{h}_l^T + \mathbf{e} \mathbf{h}_l^T
\label{Eqn_pp}    
\end{align}
Since $\mathbf{w}\mathbf{h}_l^T$ is equally likely to be zero or one, in (\ref{Eqn_pp}) $r_l$ is equally likely to be zero or one.
Using this we now prove that $P(\mathbf{y}_j,f_1) = P(\mathbf{y}_j,f_2)$, where $f_1(X)$ and $f_2(X)$ are any two factors of $X^n+1$.
For any factor $f(X)$ of $X^n+1$, $P(\mathbf{y},f)$ is given by,
\begin{equation}
\begin{aligned}
P(\mathbf{y},f) &= \frac{1}{\deg(f)} \sum_{l=0}^{\deg(f)-1} \mathbb{P}[r_l = 0] \\
		    &\stackrel{(a)}{=} \frac{1}{\deg(f)} \sum_{l=0}^{\deg(f)-1} \frac{1}{2} \\
		    &= \frac{1}{2} \frac{1}{\deg(f)} \sum_{l=0}^{\deg(f)-1} 1 
		    = \frac{1}{2}, 		    
\end{aligned}
\label{Eqn_chinese_Pjf_expression_appendix}
\end{equation}
where the equality in $(a)$ is obtained since each $r_l$ is equally likely to be zero or one. 
It can be seen that the value of $P(\mathbf{y},f)$ does not depend on the chosen $f(X)$.
This implies that $P(\mathbf{y},f_1) = P(\mathbf{y},f_2)$ and the proof is complete.
\hfill $\blacksquare$



\bibliographystyle{IEEEtran}
\bibliography{idcode_April16}

\begin{thebibliography}{10}
\providecommand{\url}[1]{#1}
\csname url@samestyle\endcsname
\providecommand{\newblock}{\relax}
\providecommand{\bibinfo}[2]{#2}
\providecommand{\BIBentrySTDinterwordspacing}{\spaceskip=0pt\relax}
\providecommand{\BIBentryALTinterwordstretchfactor}{4}
\providecommand{\BIBentryALTinterwordspacing}{\spaceskip=\fontdimen2\font plus
\BIBentryALTinterwordstretchfactor\fontdimen3\font minus
  \fontdimen4\font\relax}
\providecommand{\BIBforeignlanguage}[2]{{%
\expandafter\ifx\csname l@#1\endcsname\relax
\typeout{** WARNING: IEEEtran.bst: No hyphenation pattern has been}%
\typeout{** loaded for the language `#1'. Using the pattern for}%
\typeout{** the default language instead.}%
\else
\language=\csname l@#1\endcsname
\fi
#2}}
\providecommand{\BIBdecl}{\relax}
\BIBdecl

\bibitem{LinCostello2004}
S.~Lin and D.~Costello, \emph{Error Control Coding}, 2nd~ed.\hskip 1em plus
  0.5em minus 0.4em\relax Englewood Cliffs, New Jersey, USA: Prentice-Hall,
  2004.

\bibitem{Rice95}
B.~Rice, ``Determining the parameters of a rate 1/n convolutional encoder over
  {GF}(q),'' in \emph{Proceedings of 3rd International Conference on Finite
  Fields and Applications}, Glasgow, Scotland, July 1995.

\bibitem{Planquette96}
G.~Planquette, ``Identification de trains binaires cod{\'e}s,'' \emph{Ph.D.
  Thesis, Universite de Rennes I, France}, 1996.

\bibitem{Filiol97}
E.~Filiol, ``Reconstruction of convolutional encoders over {GF}(q),'' in
  \emph{Crytography and Coding: Lecture Notes in Computer Science}, vol. 1335,
  Berlin, Heidelberg, 1997, pp. 101--109.

\bibitem{Valembois2001}
A.~Valembois, ``Detection and recognition of a binary linear code,''
  \emph{Discrete Applied Mathematics}, vol. 111, pp. 199--218, July 2001.

\bibitem{Marazin2011}
M.~Marazin, R.~Gautier, and G.~Burel, ``Blind recovery of $k/n$ rate
  convolutional encoders in a noisy environment,'' \emph{EURASIP Journal on
  Wireless Communications and Networking}, no.~1, pp. 1--9, 2011.

\bibitem{DingelHau2007}
J.~Dingel and J.~Hagenauer, ``Parameter estimation of a convolutional encoder
  from noisy observations,'' in \emph{Proceedings of IEEE International
  Symposium on Information Theory}, Nice, France, June 2007, pp. 1776--1780.

\bibitem{Barbier2005}
J.~Barbier, ``Reconstruction of turbo-code encoders,'' in \emph{Proceedings of
  SPIE}, vol. 5819, 2005, pp. 463--473.

\bibitem{CoteSen2010}
M.~C{\^o}te and N.~Sendrier, ``Reconstruction of a turbo-code interleaver from
  noisy observation,'' in \emph{Proceedings of IEEE International Symposium on
  Information Theory}, Austin, Texas, 2010, pp. 2003--2007.

\bibitem{SicotHouBaJournal2009}
G.~Sicot, S.~Houcke, and J.~Barbier, ``Blind detection of interleaver
  parameters,'' \emph{Signal Processing}, vol.~89, no.~4, pp. 450--462, April
  2009.

\bibitem{CluF2009}
M.~Cluzeau and M.~Finiasz, ``Recovering a code's length and synchronization
  from a noisy intercepted bitstream,'' in \emph{Proceedings of IEEE
  International Symposium on Information Theory}, Seoul, Korea, July 2009, pp.
  2737--2741.

\bibitem{Cluzeau2006}
M.~Cluzeau, ``Block code reconstruction using iterative decoding techniques,''
  in \emph{Proceedings of IEEE International Symposium on Information Theory},
  Seattle, USA, July 2006, pp. 2269--2273.

\bibitem{Moosavi_journal}
R.~Moosavi and E.~Larsson, ``Fast blind recognition of channel codes,''
  \emph{IEEE Transactions on Communications}, vol.~62, no.~5, pp. 1393--1405,
  2014.

\bibitem{LeeSong2012_Korea}
H.~Lee, C.~Park, J.~Lee, and Y.~Song, ``Reconstruction of {BCH} codes using
  probability compensation,'' in \emph{Proceedings of IEEE APCC}, Jeju Island,
  Korea, October 2012, pp. 591--594.

\bibitem{Chabot_thesis}
C.~Chabot, ``Reconnaissance de codes, structure des codes quasi-cycliques,''
  \emph{PhD thesis, University of Limoges}, 2009.

\bibitem{EuropeanWireless2014}
A.~Yardi, S.~Vijayakumaran, and A.~Kumar, ``Blind reconstruction of binary
  cyclic codes,'' in \emph{Proceedings of European Wireless}, Barcelona, Spain,
  May 2014, pp. 849--854.

\bibitem{TCOMM_2016}
------, ``Blind reconstruction of binary cyclic codes from unsynchronized
  bitstream,'' \emph{IEEE Transactions on Communications}, vol.~64, no.~7, pp.
  2693--2706, 2016.

\bibitem{Zhou2013_Entropy_new}
J.~Zhou, Z.~Huang, S.~Su, and Y.~Shaowu, ``Blind recognition of binary cyclic
  codes,'' \emph{EURASIP Journal on Wireless Communications and Networking},
  vol. 2013, no.~1, pp. 1--17, 2013.

\bibitem{Zhou2013_Entropy}
J.~Zhou, Z.~Huang, C.~Liu, S.~Su, and Y.~Zhang,
  ``Information-dispersion-entropy-based blind recognition of binary {BCH}
  codes in soft decision situations,'' \emph{Entropy}, vol.~15, no.~5, pp.
  1705--1725, 2013.

\bibitem{Poor94}
H.~V. Poor, \emph{Introduction to Signal Detection and Estimation},
  2nd~ed.\hskip 1em plus 0.5em minus 0.4em\relax New York, USA:
  Springer-Verlag, 1994.

\bibitem{ISIT2014}
A.~Yardi, A.~Kumar, and S.~Vijayakumaran, ``Channel-code detection by a
  third-party receiver via the likelihood ratio test,'' in \emph{Proceedings of
  IEEE International Symposium on Information Theory}, Honolulu, HI, USA, June
  2014, pp. 1051--1055.

\bibitem{Macwilliams_Sloane_1977}
F.~MacWilliams and N.~Sloane, \emph{The Theory of Error Correcting
  Codes}.\hskip 1em plus 0.5em minus 0.4em\relax Amsterdam,Netherlands:
  North-Holland Publishing Company, 1977.

\bibitem{Lidl86}
R.~Lidl and H.~Niederreiter, \emph{Introduction to Finite Fields and Their
  Applications}.\hskip 1em plus 0.5em minus 0.4em\relax Cambridge, United
  Kingdom: Cambridge University Press, 1986.

\bibitem{Cancellieri_2015}
G.~Cancellieri, \emph{Polynomial Theory of Error Correcting Codes}.\hskip 1em
  plus 0.5em minus 0.4em\relax Cham, Switzerland: Springer, 2015.

\bibitem{Peterson_1996}
W.~Peterson and E.~Weldon, \emph{Error-Correcting Codes}, 2nd~ed.\hskip 1em
  plus 0.5em minus 0.4em\relax Cambridge, Massachusetts, USA: MIT Press, 1996.

\bibitem{Huffman_Pless_ECC}
W.~Huffman and V.~Pless, \emph{Fundamentals of Error-Correcting Codes}.\hskip
  1em plus 0.5em minus 0.4em\relax Cambridge, United Kingdom: Cambridge
  University Press, 2003.

\bibitem{ThomasCover2006}
T.~Cover and J.~Thomas, \emph{Elements of Information Theory}.\hskip 1em plus
  0.5em minus 0.4em\relax New York, USA: Wiley, 1991.

\bibitem{Artin_Algebra}
M.~Artin, \emph{Algebra}.\hskip 1em plus 0.5em minus 0.4em\relax New Jersey,
  USA: Prentice-Hall, 1991.

\bibitem{Sullivan67}
D.~Sullivan, ``A fundamental inequality between the probabilities of binary
  subgroups and cosets,'' \emph{IEEE Transactions on Information Theory},
  vol.~13, no.~1, pp. 91--94, 1967.

\end{thebibliography}

\end{document}